\documentclass[aps,pra,twocolumn,showpacs,showkeys]{revtex4}
\usepackage{graphicx,amsmath,amssymb,float,bm,times}
\usepackage[breaklinks]{hyperref}

\newcommand{\ek}{\epsilon_{\mathbf{k}}}

\newcommand{\Ek}{E_{\mathbf{k}}}

\newcommand{\phik}{\varphi_{\mathbf{k}}}
\newcommand{\phikq}{\varphi_{{\mathbf{k}}-{\mathbf{q}}/2}}

\newcommand{\sumk}{\sum_{\mathbf{k}}}

\newcommand{\sumq}{\sum_{\mathbf{q}}}

\newcommand{\Omegaq}{\Omega_{\mathbf{q}}}

\newcommand{\uk}{u_{\mathbf{k}}}
\newcommand{\vk}{v_{\mathbf{k}}}
\newcommand{\createa}[1]{a^\dagger_{#1}}
\newcommand{\destroya}[1]{a^{\phantom \dagger}_{#1}}
\newcommand{\createb}[1]{b^\dagger_{#1}}
\newcommand{\destroyb}[1]{b^{\phantom\dagger}_{#1}}

\DeclareGraphicsExtensions{.eps,.ps}

\begin{document}

\title{Applying BCS-BEC Crossover Theory To High Temperature
  Superconductors and Ultracold Atomic Fermi Gases}

\author{Qijin Chen$^1$, Jelena Stajic$^2$, 
and K.  Levin$^1$}

\affiliation{$^1$James Franck Institute and Department of Physics,
University of Chicago, Chicago, Illinois 60637}
\affiliation{$^2$Los Alamos National Laboratory,
Los Alamos, New Mexico, 87545}

\date{\today}

\begin{abstract}
  This review is written at the time of the twentieth anniversary of the
  discovery of high temperature superconductors, which, nearly coincides
  with the important discovery of the superfluid phases of ultracold
  trapped fermionic atoms. We show how these two subjects have much in
  common. Both have been addressed from the perspective of the BCS-Bose
  Einstein condensation (BEC) crossover scenario, which is designed to
  treat short coherence length superfluids with transition temperatures
  which are ``high", with respect to the Fermi energy.  A generalized
  mean field treatment of BCS-BEC crossover at general temperatures $T$,
  based on the BCS-Leggett ground state, has met with remarkable success
  in the fermionic atomic systems. Here we summarize this success in the
  context of four different cold atom experiments, all of which provide
  indications, direct or indirect, for the existence of a pseudogap.
  This scenario also provides a physical picture of the pseudogap phase
  in the underdoped cuprates which is a central focus of high $T_c$
  research.  We summarize successful applications of BCS-BEC crossover
  to key experiments in high $T_c$ systems including the phase diagram,
  specific heat, and vortex core STM data, along with the Nernst effect,
  and exciting recent data on the superfluid density in very underdoped
  samples,
\end{abstract}

\maketitle
\tableofcontents

\section{\textbf{Introduction}}
\label{sec:1}

\subsection{Historical Background}
\label{sec:1anew}

Most workers in the field of high $T_c$ superconductivity would agree
that we have made enormous progress in the last 20 years in
characterizing these materials and in identifying key theoretical
questions and constructs.  Experimental progress, in large part, comes
from transport studies \cite{Timusk,LoramPhysicaC} in addition to three
powerful spectroscopies: photoemission \cite{arpesanl1,arpesstanford},
neutron
\cite{Aeppli1991,Keimer,Birgeneau,Mason,Aeppli1997,MookNature,Rossat-Mignod1991,Tranquada} and Josephson interferometry \cite{Urbana,IBM,Maryland}.
Over the last two decades, theorists have emphasized different aspects
of the data, beginning with the anomalous normal state associated with
the highest $T_c$ systems (``optimal doping") and next, establishing the
nature and implications of the superconducting phase, which was
ultimately revealed to have a $d$-wave symmetry. Now at the time of this
twenty year anniversary, one of the most exciting areas of research
involves the normal state again, but in the low $T_c$ regime, where the
system is ``underdoped" and in proximity to the Mott insulating phase.
We refer to this unusual phase as the ``pseudogap state".

This pseudogap phase represents a highly anomalous form of
superconductivity in the sense that there is an excitation gap present
at the superfluid transition temperature $T_c$ where long range order
sets in.  The community has struggled with two generic classes of
scenarios for explaining the pseudogap and its implications below $T_c$.
Either the excitation gap is intimately connected to the superconducting
order reflecting, for example, the existence of ``pre-formed pairs", or
it is extrinsic and associated with a competing ordered state unrelated
to superconductivity.

The emphasis of this Review is on the pseudogap state as addressed by a
particular preformed pair scenario which has its genesis in what is now
referred to as ``BEC-Bose Einstein condensation (BEC) crossover theory".
Here one contemplates that the attraction (of unspecified origin) which
leads to superconductivity is stronger than in conventional
superconductivity.  In this way fermion pairs form before they Bose
condense, much as in a Bose superfluid.  In support of this viewpoint
for the cuprates are the observations that: (i) the coherence length
$\xi$ for superconductivity is anomalously short, around $10 $\AA\ as
compared with $1000$\AA\ for a typical superconductor.  Moreover (ii)
the transition temperatures are anomalously high, and (iii) the systems
are close to two dimensional (2D) (where pre-formed pair or fluctuation
effects are expected to be important). Finally, (iv) the pseudogap has
the same $d$-wave symmetry \cite{footnoteongapsymmetry} as the
superconducting order parameter \cite{arpesstanford,arpesanl1} and there
seems to be a smooth evolution of the excitation gap from above to below
$T_c$.

To investigate this BCS-BEC crossover scenario we have the particular
good fortune today of having a new class of atomic physics experiments
involving ultracold trapped fermions which, in the presence of an
applied magnetic field, have been found to have a continuously tunable
attractive interaction. At high fields the system exhibits BCS-like
superfluidity, whereas at low fields one sees BEC-like behavior.

This Review presents a consolidated study of both the pseudogap phase of
the cuprates and recent developments in ultracold fermionic superfluids.
The emphasis of these cold atom experiments is on the so-called unitary
or strong scattering regime, which is between the BEC and BCS limits,
but on the fermionic side.  The superfluid state in this intermediate
regime is also referred to in the literature as a ``resonant superfluid"
\cite{Holland,Timmermans}. Here we prefer to describe it as the
``pseudogap phase", since that is more descriptive of the physics and
underlines the close analogy with high $T_c$ systems. Throughout this
Review we will use these three descriptive phrases interchangeably.

\subsection{Fermionic Pseudogaps and Meta-stable
Pairs: Two Sides of the Same Coin}
\label{sec:1a}

BCS-BEC crossover theory is based on the observations of Eagles
\cite{Eagles} and Leggett \cite{Leggett} who independently noted that
the BCS ground state wavefunction
\begin{equation}
  \Psi_0=\Pi_{\bf k}(\uk+\vk c_k^{\dagger} c_{-k}^{\dagger})|0\rangle  
\label{eq:1a}
\end{equation}
had a greater applicability than had been appreciated at the time of its
original proposal by Bardeen, Cooper and Schrieffer (BCS).  As the
strength of the attractive pairing interaction $U$ ($<0$) between
fermions is increased, this wavefunction is also capable of describing a
continuous evolution from BCS like behavior to a form of BEC. What is
essential is that the chemical potential $\mu$ of the fermions be self
consistently computed as $U$ varies.

The variational parameters $\vk$ and $u_{\bf k}$ are usually represented
by the two more directly accessible parameters $\Delta_{sc}(0)$ and
$\mu$, which characterize the fermionic system. Here $\Delta_{sc}(0)$ is
the zero temperature superconducting order parameter. These fermionic
parameters are uniquely determined in terms of $U$ and the fermionic
density $n$. The variationally determined self consistency conditions
are given by two BCS-like equations which we refer to as the ``gap" and
``number" equations respectively.
\begin{eqnarray}
\Delta_{sc}(0)&=&-U \sum_{\bf k} \Delta_{sc}(0) \frac{1}{2 \Ek} \nonumber  \\
n&=&2 \sum_{\bf k} \left[ 1 -\frac{\ek - \mu}{\Ek} 
 \right]  
\label{eq:2}
\end{eqnarray}
where 
\begin{equation}
\Ek \equiv \sqrt{ (\ek -\mu)^2 + \Delta_{sc}^2 (0) }
\label{eq:dispersion}
\end{equation}
and $\ek=\hbar^2k^2/2 m$ are the dispersion relations for the Bogoliubov
quasiparticles and free fermions, respectively.  An additional advantage
of this formalism is that Bogoliubov de Gennes theory, a real space
implementation of this ground state, can be used to address the effects
of inhomogeneity and external fields at $T=0$.  This has been widely
used in the crossover literature.

Within this ground state there have been extensive studies \cite{Micnas}
of collective modes \cite{randeriareview,Cote} and effects of two
dimensionality \cite{randeriareview}.  Nozieres and Schmitt-Rink were
the first \cite{NSR} to address non-zero $T$. We will outline some of
their conclusions later in this Review.  Randeria and co-workers
reformulated the approach of Nozieres and Schmitt-Rink (NSR) and
moreover, raised the interesting possibility that crossover physics
might be relevant to high temperature superconductors
\cite{randeriareview}.  Subsequently other workers have applied this
picture to the high $T_c$ cuprates \cite{Chen2,Micnas1,Ranninger} and
ultracold fermions \cite{Holland,Timmermans,Milstein,Griffin} as well as
formulated alternative schemes \cite{Griffin2,Strinati2} for addressing
$ T \neq 0$.  Importantly, a number of experimentalists, most notably
Uemura \cite{Uemura}, have claimed evidence in support
\cite{Renner,Deutscher,Junod} of the BCS-BEC crossover picture for high
$T_c$ materials.

Compared to work on the ground state, considerably less has been written
on crossover effects at non-zero temperature based on Eq.~(\ref{eq:1a}).
Because our understanding has increased substantially since the
pioneering work of NSR, and because they are the most interesting, this
review is focused on these finite $T$ effects.

The importance of obtaining a generalization of BCS theory which
addresses the crossover from BCS to BEC ground state at general
temperatures $ T \leq T_c$ cannot be overestimated.  BCS theory as
originally postulated can be viewed as a paradigm among theories of
condensed matter systems; it is complete, in many ways generic and model
independent, and well verified experimentally.  The observation that the
wavefunction of Eq.~(\ref{eq:1a}) goes beyond strict BCS theory,
suggests that there is a larger mean field theory to be addressed.
Equally exciting is the possibility that this mean field theory can be
discovered and simultaneously tested in a very controlled fashion using
ultracold fermionic atoms \cite{Holland,Timmermans}.  Mean field
approaches are always approximate.  We can ascribe the simplicity and
precision of BCS theory to the fact that in conventional superconductors
the coherence length $\xi$ is extremely long. As a result, the kind of
averaging procedure implicit in mean field theory becomes nearly exact.
Once $\xi$ becomes small BCS is not expected to work at the same level
of precision.  Nevertheless even when they are not exact, mean field
approaches are excellent ways of building up intuition.  And further
progress is not likely to be made without investigating first the
simplest of mean field approaches, associated with Eq.~(\ref{eq:1a}).

\begin{figure}
\includegraphics[width=2.3in,clip]{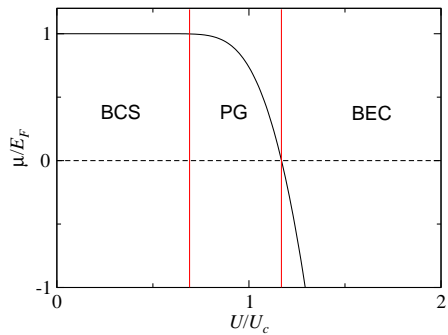}
\caption{Behavior of the $T=0$ chemical potential $\mu$ in the three
  regimes. $\mu$ is essentially pinned at the Fermi temperature $E_F$ in
  the BCS regime, whereas it becomes negative in the BEC regime. The PG
  (pseudogap) case corresponds to non-Fermi liquid based
  superconductivity in the intermediate regime.}
\label{fig:2a}
\end{figure}

The effects of BEC-BCS crossover are most directly reflected in the
behavior of the fermionic chemical potential $\mu$.  We plot the
behavior of $\mu$ in Fig.~\ref{fig:2a}, which indicates the BCS and
BEC regimes.  In the weak coupling regime $\mu = E_F$ and ordinary BCS
theory results.  However at sufficiently strong coupling, $\mu$ begins
to decrease, eventually crossing zero and then ultimately becoming
negative in the BEC regime, with increasing $|U|$.  We generally view $\mu
= 0 $ as a crossing point.  For positive $\mu$ the system has a remnant
of a Fermi surface, and we say that it is ``fermionic".  For negative
$\mu$, the Fermi surface is gone and the material is ``bosonic".

\textit{The new and largely unexplored physics of this problem lies in
  the fact that once outside the BCS regime, but before BEC,
  superconductivity or superfluidity emerge out of a very exotic,
  non-Fermi liquid normal state}.  Emphasized in Fig.~\ref{fig:2a} is
this intermediate (i.e., pseudogap or PG) regime having positive $\mu$
which we associate with non-Fermi liquid based superconductivity
\cite{Chen2,Chen3,Maly1}.  Here, the onset of superconductivity occurs
in the presence of fermion pairs. Unlike their counterparts in the BEC
limit, these pairs are not infinitely long lived.  Their presence is
apparent even in the normal state where an energy must be applied to
create fermionic excitations.  This energy cost derives from the
breaking of the metastable pairs.  Thus we say that there is a
``pseudogap" (PG) at and above $T_c$. It will be stressed throughout
this Review that gaps in the fermionic spectrum and bosonic degrees of
freedom are two sides of the same coin.  A particularly important
observation to make is that the starting point for crossover physics is
based on the fermionic degrees of freedom.  A non-zero value of the
excitation gap $\Delta$ is equivalent to the presence of metastable or
stable fermion pairs. And it is only in this indirect fashion that we
can probe the presence of these ``bosons", within the framework of
Eq.~(\ref{eq:1a}).

\begin{figure}
\includegraphics[width=2.3in,clip]{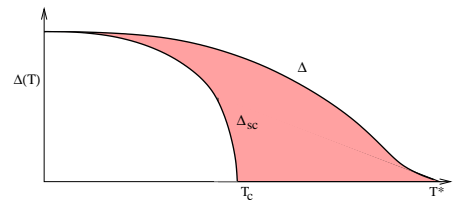}
\caption{Contrasting behavior of the excitation gap $\Delta(T)$ and
  order parameter $\Delta_{sc}(T)$ versus temperature in the pseudogap
  regime. The height of the shaded region reflects the number of
  noncondensed pairs, at each temperature.}
\label{fig:Delta_Deltasc}
\end{figure}

In many ways this crossover theory appears to represent a more generic
form of superfluidity.  Without doing any calculations we can anticipate
some of the effects of finite temperature. Except for very weak
coupling, \textit{pairs form and condense at different temperatures}.
More generally, in the presence of a moderately strong attractive
interaction it pays energetically to take some advantage and to form
pairs (say roughly at temperature $T^*$) within the normal state. Then,
for statistical reasons these bosonic degrees of freedom ultimately are
driven to condense at $T_c < T^*$, as in BEC.

Just as there is a distinction between $T_c$ and $T^*$, \textit{there
  must be a distinction between the superconducting order parameter
  $\Delta_{sc}$ and the excitation gap $\Delta$}.  In
Fig.~\ref{fig:Delta_Deltasc} we present a schematic plot of these two
energy parameters.  It may be seen that the order parameter vanishes at
$T_c$, as in a second order phase transition, while the excitation gap
turns on smoothly below $T^*$.  It should also be stressed that there is
only one gap energy scale in the ground state \cite{Leggett} of
Eq.~(\ref{eq:1a}).  Thus $\Delta_{sc}(0) = \Delta(0)$.

\begin{figure}
\includegraphics[width=3.4in,clip]{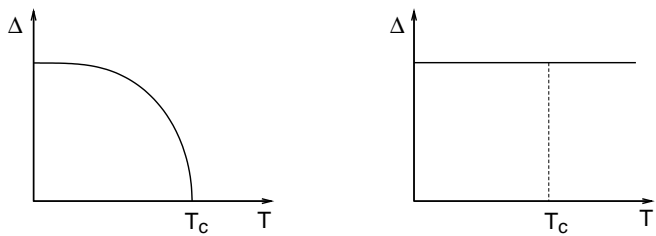}
\caption{Comparison of temperature dependence of excitation gaps in BCS
  (left) and BEC (right) limits. The gap vanishes at $T_c$ for the
  former while it is essentially $T$-independent for the latter.} 
\label{fig:2}
\end{figure}

In addition to the distinction between $\Delta$ and $\Delta_{sc}$,
another important way in which \textit{bosonic degrees of freedom are
  revealed is indirectly through the temperature dependence of $\Delta$.
  In the BEC regime where fermionic pairs are pre-formed, $\Delta$ is
  essentially constant for all $ T \leq T_c$ (as is $\mu$)}. By contrast
in the BCS regime it exhibits the well known temperature dependence of
the superconducting order parameter.  This behavior is illustrated in
Fig.~\ref{fig:2}.

\begin{figure}
\centerline{\Large{ \mbox{\hspace{0.02in} \emph{\small{BCS}} \hspace{0.6in}
\emph{\small{Pseudogap (PG)}} \hspace{0.5in} \emph{\small{BEC}}} }}
 \vskip 0.05in \centerline{\includegraphics[bb = 0 0 510 150,
 width=3.4in, clip]{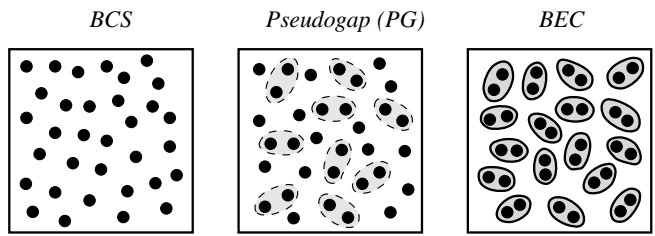}}
\caption{The character of the excitations in the BCS-BEC crossover both
  above and below $T_c$. The excitations are primarily fermionic
  Bogoliubov quasiparticles in the BCS limit and bosonic pairs (or
  ``Feshbach bosons") in the BEC limit.  For atomic Fermi gases, the
  ``virtual molecules" in the PG case consist primarily of ``Cooper''
  pairs of fermionic atoms.  }
\label{fig:3}
\end{figure}

Again, without doing any calculations we can make one more inference
about the nature of crossover physics at finite $T$. \textit{The
  excitations of the system must smoothly evolve from fermionic in the
  BCS regime to bosonic in the BEC regime}.  In the intermediate case,
the excitations are a mix of fermions and meta-stable pairs. Figure
\ref{fig:3} characterizes the excitations out of the condensate as well
as in the normal phase. This schematic figure will play an important
role in our thinking throughout this Review.

\subsection{Introduction to high $T_c$ Superconductivity: Pseudogap
Effects}
\label{sec:1B}

This Review deals with the intersection of two fields and two important
problems: high temperature superconductors and ultracold fermionic atoms
in which, through Feshbach resonance effects, the attractive interaction
may be arbitrarily tuned by a magnetic field.  Our focus is on the
broken symmetry phase and how it evolves from the well known ground
state at $T=0$ to $T=T_c$. We begin with a brief overview
 \cite{Timusk,LoramPhysicaC} of pseudogap effects in high temperature
superconductors.  A study of concrete data in these systems provides a
rather natural way of building intuition about non-Fermi liquid based
superfluidity, and this should, in turn, be useful for the cold atom
community.

It has been argued by some \cite{Chen1,Micnas1,Ranninger,Strinati3,YY}
that a BCS-BEC crossover-induced pseudogap is the origin of the
mysterious normal state gap observed in high temperature
superconductors.  While this is a highly contentious subject some of the
arguments in favor of this viewpoint (beyond those listed in Section
\ref{sec:1anew}) rest on the following observations: (i) To a good
approximation the pseudogap onset temperature \cite{Fischer2,Oda} $ T^*
\approx 2 \Delta(0)/4.3 $ which satisfies the BCS scaling relation. (ii)
There is widespread evidence for pseudogap effects both above
\cite{Timusk,LoramPhysicaC} as well as (iii) below \cite{Loram,JS1}
$T_c$.  (iv) In addition, it has also been argued that short coherence
length superconductors may quite generally exhibit a distinctive form of
superconductivity \cite{Uemura} which sets them apart from conventional
superconductors.  One might want, then, to concentrate on this more
generic feature, (rather than on more exotic aspects), which they have
in common with other superconductors in their distinctive class.

\begin{figure}
\includegraphics[width=2.3in,clip]{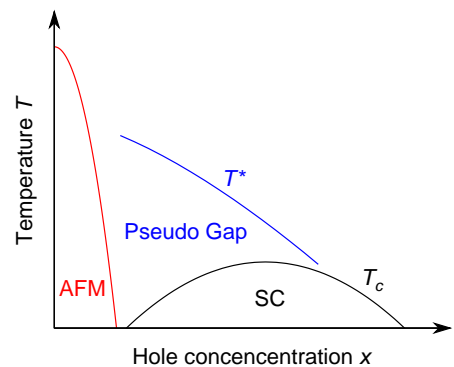}
\caption{Typical phase diagram of hole-doped high $T_c$
superconductors.
There exists a pseudogap phase above $T_c$ in the underdoped regime.
Here SC denotes superconductor, and $T^*$ is the temperature at which
the pseudogap smoothly turns on.}
\label{fig:5}
\end{figure}

\begin{figure}
\centerline{\includegraphics[width=2.5in]{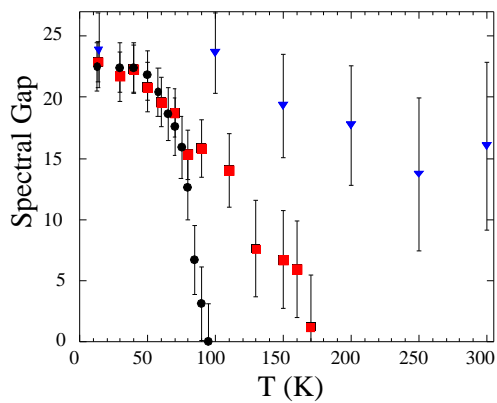}}
\caption{Temperature dependence of the excitation gap at the antinodal
  point ($\pi$,0) in Bi$_2$Sr$_2$CaCu$_2$O$_{8+\delta}$ (BSCCO) for
  three different doping concentrations from near-optimal (discs) to
  heavy underdoping (inverted triangles), as measured by angle-resolved
  photoemission spectroscopy (from Ref.~\onlinecite{arpesanl1}).}
\label{fig:7}
\end{figure}

\begin{figure*}[t]
\includegraphics[width=4.5in]{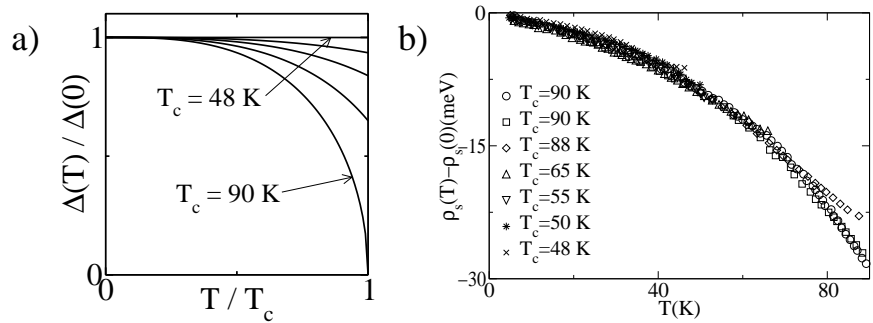}
\caption{Temperature dependence of fermionic excitation gaps $\Delta$
and superfluid density $\rho_s$ for various doping concentrations
(from Ref.~\onlinecite{JS1}). When $\Delta(T_c) \neq 0$, there is little
correlation between $\Delta(T)$ and $\rho_s(T)$; this figure suggests
that something other than fermionic quasi-particles (\textit{e.g.},
bosonic excitations) may be responsible for the disappearance of
superconductivity with increasing $T$.  Figure (b) shows a
quasi-universal behavior for the slope $d\rho_s/dT$ at different
doping concentrations, despite the highly non-universal behavior for
$\Delta(T)$.}
\label{fig:10}
\end{figure*}

In Fig.~\ref{fig:5} we show a sketch of the phase diagram for the
hole-doped copper oxide superconductors. Here $x$ represents the
concentration of holes which can be controlled by, say, adding Sr
substitutionally to La$_{1-x}$Sr$_{x}$CuO$_4$. At zero and small $x$ the
system is an antiferromagnetic (AFM) insulator. Precisely at half
filling ($x=0$) we understand this insulator to derive from Mott
effects. These Mott effects may or may not be the source of the other
exotic phases indicated in the diagram, i.e., the superconducting (SC)
and the ``pseudogap'' (PG) phases.  Once AFM order disappears the system
remains insulating until a critical hole concentration (typically around
a few percent) when an insulator-superconductor transition is
encountered. Here photoemission studies \cite{arpesstanford,arpesanl1}
suggest that once this line is crossed, $\mu$ appears to be positive.
For $ x \leq 0.2$, the superconducting phase has a non-Fermi liquid (or
pseudogapped) normal state \cite{LoramPhysicaC}.  We note an important
aspect of this phase diagram at low $x$. As the pseudogap becomes
stronger ($T^*$ increases), superconductivity as reflected in the
magnitude of $T_c$ becomes weaker.

Figure \ref{fig:7} indicates the temperature dependence of the
excitation gap for three different hole stoichiometries. These data
\cite{arpesanl1} were taken from angle resolved photoemission
spectroscopy (ARPES) measurement. For one sample shown as circles,
(corresponding roughly to ``optimal" doping) the gap vanishes roughly at
$T_c$ as might be expected for a BCS superconductor. At the other
extreme are the data indicated by inverted triangles in which an
excitation gap appears to be present up to room temperature, with very
little temperature dependence.  This is what is referred to as a highly
underdoped sample (small $x$), which from the phase diagram can be seen
to have a rather low $T_c$.  Moreover, $T_c$ is not evident
in these data on underdoped samples.

While the high $T_c$ community has focused on pseudogap effects above
$T_c$, there is a good case to be made that these effects also persist
below.  STM data \cite{Renner} taken below $T_c$ within a vortex core
indicate that there is a clear depletion of the density of states around
the Fermi energy in the normal phase within the core.  These data
underline the fact that the existence of an energy gap has little or
nothing to do with the existence of phase coherent superconductivity. It
also underlines the fact that pseudogap effects effectively persist
below $T_c$; the normal phase underlying superconductivity for $ T \leq
T_c$ is not a Fermi liquid.

Analysis of thermodynamical data \cite{Loram,LoramPhysicaC} has led to a
similar inference.  For the PG case, the entropy extrapolated into the
superfluid phase, based on Fermi liquid theory, becomes negative. In
this way Loram and co-workers \cite{Loram} deduced that the normal phase
underlying the superconducting state is not a Fermi liquid.  Rather,
they claimed to obtain proper thermodynamics, it must be assumed that
this state contains a persistent pseudogap.  In this way they argued for
a distinction between the excitation gap $\Delta$ and the
superconducting order parameter, within the superconducting phase.  To
fit their data they presume a modified fermionic dispersion $ \Ek =
\sqrt{ (\ek -\mu)^2 + \Delta^2 (T) }$ where
\begin{equation}
\Delta^2(T) = \Delta_{sc}^2(T) + \Delta_{pg}^2
\label{eq:2a}
\end{equation}
Here $\Delta_{pg}$ is taken on phenomenological grounds to be
$T$-independent.  While Eq.~(\ref{eq:2a}) is also found
in BCS-BEC crossover theory, there are important differences. In
the latter approach 
$\Delta_{pg} \rightarrow
0$ as $ T \rightarrow 0$.

Finally, Fig.~\ref{fig:10} makes the claim for a persistent pseudogap
below $T_c$ in an even more suggestive way. Figure \ref{fig:10}a
represents a schematic plot of excitation gap data such as are shown in
Fig.~\ref{fig:7}.  Here the focus is on temperatures below $T_c$. Most
importantly, this figure indicates that the $T$ dependence in $\Delta$
varies dramatically as the stoichiometry changes. Thus, in the extreme
underdoped regime, where PG effects are most intense, there is very
little $T$ dependence in $\Delta$ below $T_c$.  By contrast at high $x$,
when PG effects are less important, the behavior of $\Delta$ follows
that of BCS theory.  What is most impressive however, is that these wide
variations in $\Delta(T)$ are \textit{not} reflected in the superfluid
density $\rho_s(T)$.  Figure \ref{fig:10} then indicates that,
\textit{despite the highly non-universal behavior for $\Delta(T)$, the
  superfluid density does not make large excursions from its BCS-
  predicted form}. This is difficult to understand if the fermionic
degrees of freedom through $\Delta(T)$ are dominating at all $x$. Rather
this figure suggests that something other than fermionic excitations is
responsible for the disappearance of superconductivity, particularly in
the regime where $\Delta(T)$ is relatively constant in $T$. At the very
least pseudogap effects must persist below $T_c$.

The phase diagram also suggests that pseudogap effects become stronger
with underdoping. How does one accommodate this in the BCS-BEC crossover
scenario? At the simplest level one may argue that as the system
approaches the Mott insulating limit, fermions are less mobile and the
effectiveness of the attraction increases.  In making the connection
between the strength of the attraction and the variable $x$ in the
cuprate phase diagram we will argue that it is appropriate to simply fit
$T^*(x)$.  In this Review we do not emphasize Mott physics because it is
not particularly relevant to the atomic physics problem.  It also seems
to be complementary to the BCS-BEC crossover scenario. It is understood
that both components are important in high $T_c$ superconductivity.  It
should be stressed that hole concentration $x$ in the cuprates plays the
role of applied magnetic field in the cold atom system. These are the
external parameters which serve to tune the BCS-BEC crossover.

Is there any evidence for bosonic degrees of freedom in the normal state
of high $T_c$ superconductors? The answer is unequivocally yes:
\textit{meta-stable bosons are observable as superconducting
  fluctuations}.  These effects are enhanced in the presence of the
quasi-two dimensional lattice structure of these materials.  In the
underdoped case, one can think of $T^*$ as marking the onset of
preformed pairs which are closely related to fluctuations of
conventional superconductivity theory, but which are made more robust as
a result of BCS-BEC crossover effects, that is, stronger pairing
attraction.  A number of people have argued \cite{Nernst,Corson1999}
that fluctuating normal state vortices are responsible for the anomalous
transport behavior of the pseudogap regime.  It has been proposed
\cite{Tan} that these data may alternatively be interpreted as
suggesting that bosonic degrees of freedom are present in the normal
state.

\subsection{Summary of Cold Atom Experiments: Crossover in the
Presence of Feshbach Resonances}
\label{sec:exp_cold_summary}

There has been an exciting string of developments over the past few
years in studies of ultracold fermionic atoms, in particular, $^6{\rm
  Li}$ and $^{40}{\rm K}$, which have been trapped and cooled via
magnetic and optical means.  Typically these traps contain $10^5$ atoms
at very low densities $\approx 10^{13}$ ${\rm cm}^{-3}$. Here the Fermi
temperature in a trap can be estimated to be of the order of a
micro-Kelvin.  It was argued on the basis of BCS theory alone
\cite{Houbiers}, and rather early on (1997), that the temperatures
associated with the superfluid phases may be attainable in these trapped
gases.  This set off a search for the ``holy grail" of fermionic
superfluidity.  That a Fermi degenerate state could be reached at all is
itself quite remarkable; this was was first reported \cite{Jin} by Jin
and deMarco in 1999.  By late 2002 reports of unusual hydrodynamics in a
degenerate Fermi gas indicated that strong interactions were present
\cite{ohara}.  This strongly interacting Fermi gas (associated with the
unitary scattering regime) has attracted widespread attention
independent of the search for superfluidity, because it appears to be a
prototype for analogous systems in nuclear physics
\cite{Baker,Heiselberg3} and in quark-gluon plasmas
\cite{Itakura,Heinz}.  Moreover, there has been a fairly extensive body
of analytic work on the ground state properties of this regime
\cite{Heiselberg2, Carlson3}, which goes beyond the simple mean field
wave function ansatz.

\begin{figure}
\includegraphics[width=3.1in,clip]{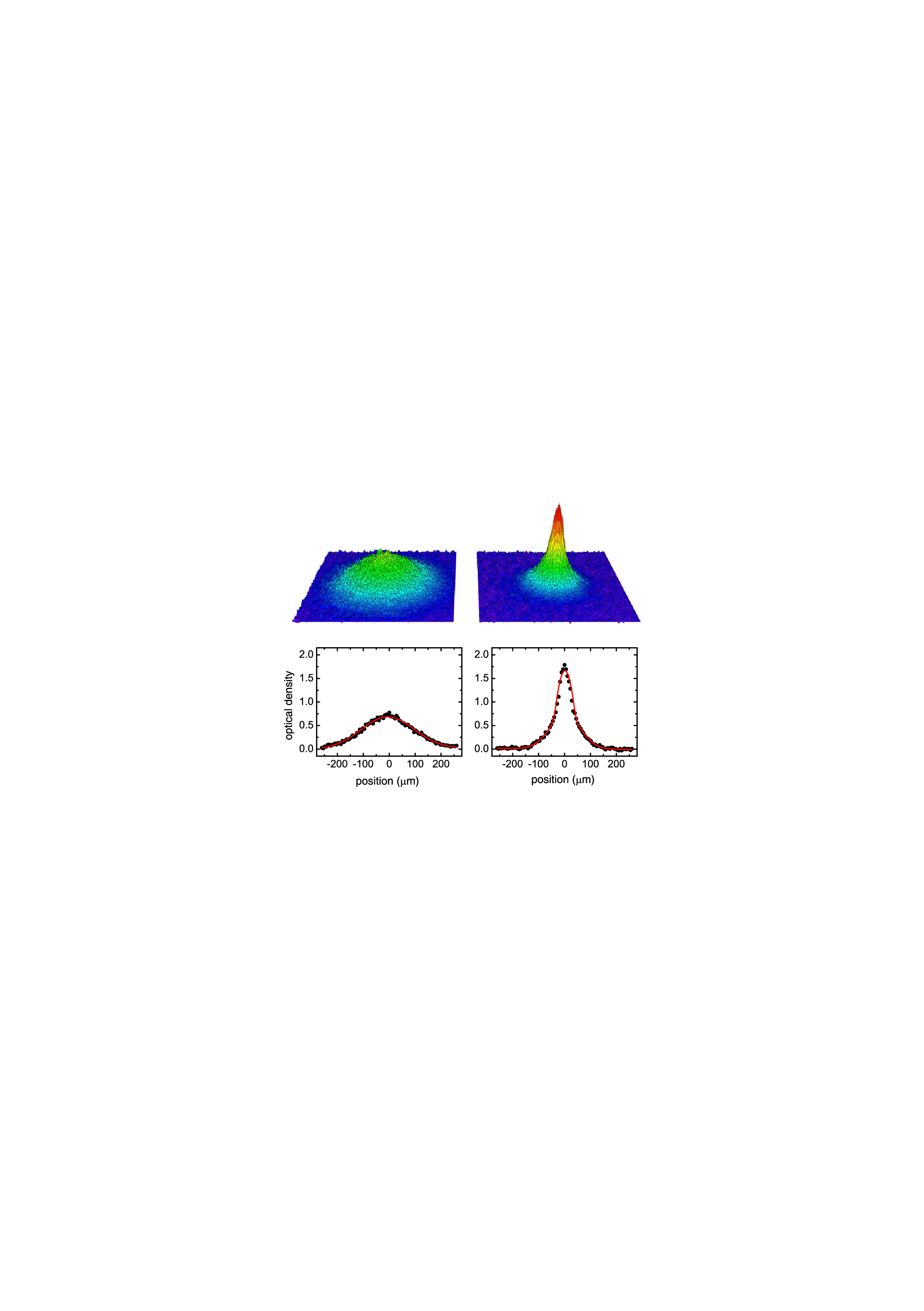}
\caption{(Color) Spatial density profiles of a molecular cloud of
  trapped $^{40}$K atoms in the BEC regime in the transverse directions
  after 20 ms of free expansion (from Ref.~\onlinecite{Jin3}), showing
  thermal molecular cloud above $T_c$ (left) and a molecular condensate
  (right) below $T_c$. (a) shows the surface plots, and (b) shows the
  cross-sections through images (dots) with bimodal fits (lines).}
\label{fig:jin}
\end{figure}

As a consequence of attractive $s$-wave interactions between fermionic
atoms in different hyperfine states, it was anticipated that dimers
could also be made. Indeed, these molecules formed rather efficiently
\cite{Regal,Hulet3,Cubizolles} as reported in mid-2003 either via three
body recombination \cite{Jochim} or by sweeping the magnetic field
across a Feshbach resonance. Moreover, they are extremely long lived
\cite{Hulet3}. From this work it was relatively straightforward to
anticipate that a Bose condensate would also be achieved.  Credit goes
to theorists such as Holland \cite{Holland} and to Griffin
\cite{Griffin} and their co-workers for recognizing that the
superfluidity need not be only associated with condensation of long
lived bosons, but in fact could also derive, as in BCS, from fermion
pairs.  In this way, it was argued that a suitable tuning of the
attractive interaction via Feshbach resonance effects, would lead to a
realization of a BCS-BEC crossover.

By late 2003 to early 2004, four groups
\cite{Jin3,Grimm,Ketterle2,Salomon3} had observed the ``condensation of
weakly bound molecules" (that is, on the $a_s > 0 $ side of resonance),
and shortly thereafter a number had also reported evidence for
superfluidity on the BCS side \cite{Jin4,Ketterle3,Thomas2,Grimm4}.  The
BEC side is the more straightforward since the presence of the
superfluid is reflected in a bi-modal distribution in the density
profile.  This is shown in Fig.~\ref{fig:jin} from
Ref.~\onlinecite{Jin3}, and is conceptually similar to the behavior for
condensed Bose atoms \cite{RMP}.  On the BEC side but near resonance,
the estimated $T_c$ is about $0.3T_F$, with condensate fractions varying
from 20\% or so to nearly 100\%. The condensate lifetimes are relatively
long in the vicinity of resonance, and fall off rapidly as one goes
deeper into the BEC.  However, for $a_s < 0$ there is no clear
expectation that the density profile will provide a signature of the
superfluid phase.

These claims that superfluidity may have been achieved on the BCS side
($a_s < 0$) of resonance were viewed as particularly exciting. The
atomic community, for the most part, felt the previous counterpart
observations on the BEC side were expected and not significantly
different from condensation in Bose atoms. The evidence for this new
form of ``fermionic superfluidity" rests on studies
\cite{Jin4,Ketterle3} that perform fast sweeps from negative $a_s$ to
positive $a_s$ across the resonance.  The field sweeps allow, in
principle, a pairwise projection of fermionic atoms (on the BCS side)
onto molecules (on the BEC side).  It is presumed that in this way one
measures the momentum distribution of fermion pairs. The existence of a
condensate was thus inferred. Other experiments which sweep across the
Feshbach resonance adiabatically, measure the size of the cloud after
release \cite{Salomon3} or within a trap \cite{Grimm2}.

Evidence for superfluidity on the BCS side, which does not rely on the
sweep experiments, has also been deduced from collective excitations of
a fermionic gas \cite{Thomas2,Grimm3}. Pairing gap measurements with
radio frequency (RF) spectroscopy probes \cite{Grimm4} have similarly
been interpreted \cite{Torma2} as providing support for the existence of
superfluidity, although more directly these experiments establish the
existence of fermion pairs.  Quite recently, evidence for a phase
transition has been presented via thermodynamic measurements and
accompanying theory \cite{ThermoScience}.  The latter, like the theory
\cite{Torma2} of RF experiments \cite{Grimm4}, is based on the formalism
presented in this Review.  A most exciting and even more recent
development has been the observation of vortices \cite{KetterleV} which
appears to provide a smoking gun for the existence of the superfluid
phase.

\section{Theoretical Formalism for BCS-BEC crossover}
\label{sec:2}

\subsection{Many-body Hamiltonian and Two-body Scattering Theory} 
\label{sec:hamiltonian}

We introduce the Hamiltonian \cite{Holland,Griffin,ourreview} used in the cold
atom and high $T_c$ crossover studies. The most general form for this
Hamiltonian consists of two types of interaction effects: those
associated with the direct interaction between fermions parametrized by
$U$, and those associated with ``fermion-boson" interactions, whose
strength is governed by $g$.
\begin{eqnarray}
\label{hamiltonian}
H&-&\mu N=\sum_{{\bf k},\sigma}(\epsilon_{\bf k}-\mu)\createa{{\bf
    k},\sigma}\destroya{{\bf k},\sigma}+\sum_{\bf q}(\epsilon_{\bf
    q}^{mb}+\nu-2 \mu)\createb{\bf q}\destroyb{\bf q}\nonumber\\ 
&+&\sum_{{\bf q},{\bf k},{\bf k'}}U({\bf k},{\bf k'})\createa{{\bf
    q}/2+{\bf k},\uparrow}\createa{{\bf q}/2-{\bf
    k},\downarrow}\destroya{{\bf q}/2-{\bf k'},\downarrow}\destroya{{\bf
    q}/2+{\bf k'},\uparrow}\nonumber\\ 
&+&\sum_{{\bf q},{\bf k}}\left(g({\bf k})\createb{{\bf q}}\destroya{{\bf
    q}/2-{\bf k},\downarrow}\destroya{{\bf q}/2+{\bf
    k},\uparrow}+h.c.\right)  
\label{eq:0c}
\end{eqnarray}
Here the fermion and boson kinetic energies are given by $\ek=k^2/2 m$,
and $\epsilon_{\bf q}^{mb}=q^2/2 M$, and $\nu$ is an important parameter
which represents the magnetic field-induced ``detuning". Here we use the
convention $\hbar=k_B=c=1$. In this two channel problem the ground state
wavefunction is slightly modified and given by
\begin{equation}
\bar{\Psi}_0 = \Psi_0 \otimes \Psi_0^B 
\end{equation}
where the molecular or Feshbach boson contribution $\Psi_0^B$ is 
as given in Ref.~\onlinecite{Ranninger2}.

Whether both forms of interactions are needed in either system is still
under debate. The bosons ($\createb{\bf k}$) of the cold atom problem
\cite{Holland,Timmermans} are referred to as belonging to the ``closed
channel".  These spin-singlet molecules represent a separate species,
not to be confused with the (``open channel") fermion pairs
($\createa{\bf k} \createa{\bf -k}$), which are associated with spin
triplet.  As a result of virtual occupation of the bound state of the
closed channel the interaction between open channel fermions can be
tuned (through applied magnetic field) to vary from weak to very strong.

In this review we will discuss the behavior of crossover physics both
with and without the closed-channel.  Previous studies of high $T_c$
superconductors have invoked a similar bosonic term
\cite{TDLee1,TDLee2,Ranninger,Larkin} as well, although less is known
about its microscopic origin.  This fermion-boson coupling is not to be
confused with the coupling between fermions and a
``pairing-mechanism"-related boson ($[b + b^\dagger]a^\dagger a$) such
as phonons in a metal superconductor. The coupling $b^\dagger a a$ and
its Hermitian conjugate represent a form of sink and source for creating
fermion pairs, in this way inducing superconductivity in some ways, as a
by-product of Bose condensation.

It is useful at this stage to introduce the $s$-wave scattering length,
$a$, defined by the low energy limit of two body scattering in vacuum.
We begin with the effects of $U$ only, presuming that $U$ is always an
attractive interaction ($U < 0$) which can be arbitrarily varied,
\begin{equation}
\frac{m}{4 \pi a } \equiv \frac{1}{U} + \sum_{\bf k} \frac{1}{2 \ek}
\label{eq:11c}
\end{equation}
We may define a critical value $U_c$ of the potential as that associated
with the binding of a two particle state in vacuum.  
We can write down an equation
for $U_c$ given by
\begin{equation}
U_c^{-1} =- \sum_{\bf k} \frac{1}{2 \ek}
\label{eq:Uc}
\end{equation}
although specific evaluation of $U_c$ requires that there be a cut-off
imposed on the above summation, associated with the range of the
potential.  \textit{The fundamental postulate of crossover theory is
  that even though the two-body scattering length changes abruptly at
  the unitary scattering condition ($|a| = \infty $), in the N-body
  problem the superconductivity varies smoothly through this point}.

Provided we redefine the appropriate ``two body" scattering length,
Equation (\ref{eq:11c}) holds even in the presence of Feshbach
effects \cite{Griffin,Milstein}.  It has been shown that $U$ in the above
equations is replaced by
\begin{equation}
U \rightarrow U_{eff} \equiv   U + \frac{g^2}{2 \mu - \nu}
\end{equation}
and we write $a \rightarrow a^*$. Experimentally, the two body scattering
length $a^*$ varies with magnetic field $B$.
Thus we have
\begin{equation}
\frac {m}{4 \pi a^*} \equiv \frac {1}{U_{eff}} + \sum_{\bf k} \frac{1}
{2 \ek}
\end{equation}
More precisely the effective interaction between two fermions is
momentum and energy dependent. It arises from a second order process
involving emission and absorption of a closed-channel molecular boson.
The net effect of the direct plus indirect interactions is given by
$U_{eff}(Q) \equiv U + g^2 D_0 (Q)$, where $D_0(Q) \equiv 1/ [
i\Omega_n- \epsilon_{\bf q}^{mb}-\nu + 2 \mu ]$ is the non-interacting
molecular boson propagator. Here and throughout we use a four-momentum
notation, $Q \equiv (\mathbf{q}, i\Omega_n)$, and its analytical
continuation, $Q \rightarrow (\mathbf{q}, \Omega+i0^+)$, and write
$\sum_Q \equiv T \sum_{\Omega_n} \sum_{\bf q}$, where $\Omega_n$ is a
Matsubara frequency.  What appears in the gap equation, however, is
$U_{eff}(Q=0)$ which we define to be $U_{eff}$.  When the open-channel
attraction $U$ is weak, clearly, $ 2\mu \leq \nu$ is required so that
the Feshbach-induced interaction is attractive.  In the extreme BEC
limit $\nu =2 \mu$.  However, when a deep bound state exists in the open
channel, such as in $^{40}$K, the system may evolve into a metastable
state such that $ 2\mu > \nu$ in the BEC regime and there is a point on
the BCS side where $U_{eff} = 0$ precisely.

\begin{figure}
\includegraphics[width=2.3in,clip]{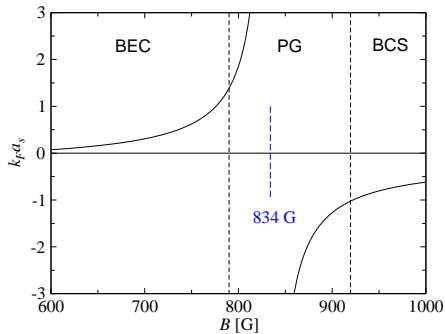}
\caption{Characteristic behavior of the scattering length for $^6$Li in
  the three regimes.}
\label{fig:9a}
\end{figure}

Figure~\ref{fig:9a} presents a plot of this scattering length $k_F a_s
\equiv k_F a^*$ for the case of $^6$Li.  It follows that $a_s$ is
negative when there is no bound state, it tends to $-\infty$ at the
onset of the bound state and to $+\infty$ just as the bound state
stabilizes.  It remains positive but decreases in value as the
interaction becomes increasingly strong. The magnitude of $a_s$ is
arbitrarily small in both the extreme BEC and BCS limits, but with
opposite sign.

\subsection{\textit{T}- Matrix-Based Approaches to BCS-BEC Crossover in the Absence of Feshbach Effects}
\label{sec:1F}

To address finite temperature in a way which is consistent with
Eq.~(\ref{eq:1a}), or with alternative ground states, one introduces a
$T$-matrix approach.  Here one solves self consistently for the single
fermion propagator ($G$) and the pair propagator ($t$).  That one stops
at this level without introducing higher order Green's functions
(involving three, and four particles, etc) is believed to be adequate
for addressing a leading order mean field theory such as that
represented by Eq.~(\ref{eq:1a}). One can see that pair-pair
(boson-boson) interactions are only treated in a (generalized) mean
field averaging procedure; they arise exclusively from the fermions and
are sufficiently weak so as not to lead to any incomplete condensation
in the ground state, as is compatible with Eq.~(\ref{eq:1a}).

In this section we demonstrate that at the $T$-matrix level there are
three distinct schemes which can be implemented to address BCS-BEC
crossover physics.  Above $T_c$, quite generally one writes for the
$t$-matrix
\begin{equation}
t(Q) = \frac{U}{1+U\chi(Q)}
\label{eq:1c}
\end{equation}
and theories differ only on what is the nature of the pair
susceptibility $\chi(Q)$, and the associated self energy of the
fermions.  Below $T_c$ one can also consider a $T$-matrix approach to
describe the particles and pairs in the condensate. For the most part we
will defer extensions to the broken symmetry phase to Section
\ref{sec:2B}.

In analogy with Gaussian fluctuations, Nozieres and Schmitt-Rink
considered \cite{NSR}
\begin{equation}
\chi_0(Q) = \sum_K G_0(K)G_0(Q-K)
\label{eq:10c}
\end{equation}
with self energy
\begin{equation}
\Sigma_0(K) = \sum_Q t(Q) G_0(Q-K)\,,
\label{eq:11d}
\end{equation}
where $G_0(K)$ is the noninteracting fermion Green's function.  The
number equation of the Nozieres Schmitt-Rink scheme
\cite{NSR,randeriareview} is then deduced in an approximate fashion
\cite{Serene} by using a leading order series for $G$ with
\begin{equation}
 G = G_0 + ~ G_0 \Sigma_0 G_0 \,.
\label{eq:13c} 
\end{equation}
It is straightforward, however, to avoid this approximation in Dyson's
equation, and a number of groups \cite{Maly1,Strinati2} have extended
NSR in this way.

Similarly one can consider 
\begin{equation}
\bar{\chi}(Q) = \sum_K G(K)G(Q-K)
\end{equation}
with self energy
\begin{equation}
\bar{\Sigma}(K) = \sum_Q t(Q) G(Q-K)\,.
\end{equation}
This latter scheme has been also extensively discussed in the
literature, by among others, Haussmann \cite{Haussmann}, Tchernyshyov
\cite{Tchern} and Yamada and Yanatse \cite{YY}.

Finally, we can contemplate the asymmetric form \cite{Chen2} for the
$T$-matrix, so that the coupled equations for $t(Q)$ and $G(K)$ are based
on
\begin{equation}
\chi(Q) = \sum_K G(K)G_0(Q-K)
\label{eq:ggo}
\end{equation}
with self energy
\begin{equation}
\Sigma(K) = \sum_Q t(Q) G_0(Q-K) \,.
\end{equation}
It should be noted, however, that this asymmetric form can be derived
from the equations of motion by truncating the infinite series at the
three particle level, $G_3$, and then factorizing the $G_3$ into one-
and two-particle Green's functions \cite{Kadanoff}. The other two
schemes are constructed diagrammatically or from a generating
functional, (as apposed to derived from the Hamiltonian).

\textit{It will be made clear in what follows that, if one's goal is to
  extend the usual crossover ground state of Eq.~(\ref{eq:1a}) to finite
  temperatures, then one must choose the asymmetric form for the pair
  susceptibility, as shown in Eq.~(\ref{eq:ggo})}.  Other approaches
such as the NSR approach to $T_c$, or that of Haussmann lead to
different ground states which should, however, be very interesting in
their own right. These will need to be characterized in future.
Indeed, the work of Strinati group has also emphasized that the ground
state associated with the $T_c$ calculations based on NSR is distinct
from that in the simple mean field theory of Eq.~(\ref{eq:1a}), and they
presented some aspects of this comparison in Ref.~\onlinecite{Strinati5}.

Other support for this $GG_0$-based $T$-matrix scheme comes from its
equivalence to self consistent Hartree-approximated Ginzburg-Landau
theory \cite{JS}.  Moreover, there have been detailed studies
\cite{Kosztin2} to demonstrate how the superfluid density $\rho_s$ can
be computed in a fully gauge invariant (Ward Identity consistent)
fashion, so that it vanishes at the self consistently determined $T_c$.
Such studies are currently missing for the case of the other two
$T$-matrix schemes.

\subsection{Extending conventional Crossover Ground State to $T \neq 0$:
  \textit{T}-matrix scheme in the presence of closed-channel molecules }
\label{sec:2B}

In the $T$-matrix scheme we employ, the pairs are described by the pair
susceptibility $\chi(Q) = \sum_K G_0(Q-K)G(K)\phikq^2$ where $G$ depends
on a BCS-like self energy $ \Sigma(K) \approx -\Delta^2 G_0(-K)\phik^2
$.  Throughout this section $\phik \equiv \exp\{-k^2/2k_0^2\}$
introduces a momentum cutoff, where $k_0$ represents the inverse range
of interaction, which is assumed infinite for a contact interaction.

The noncondensed pairs \cite{JS3} have propagator $t_{pg}(Q) =
U_{eff}(Q)/[1 + U_{eff}(Q) \chi(Q)]$, where $U_{eff}$ is the effective
pairing interaction which involves the direct two-body interaction $U$
as well as virtual excitation processes associated with the Feshbach
resonance \cite{Griffin,JS3}.  At small $Q$, $t_{pg}$ can be expanded,
after analytical continuation ($i\Omega_n \rightarrow \Omega + i0^+$),
as
\begin{equation}
t_{pg}(Q) \approx \frac { Z^{-1}}{\Omega - \Omega_q +\mu_{pair} + i \Gamma_Q}.
\label{eq:expandt}
\end{equation}
The parameters appearing in Eq.~(\ref{eq:expandt}) are discussed in more
detail in Ref. ~\onlinecite{ourreview}.  Here $Z^{-1}$ is a residue and
$ \Omega_q = q^2/2M^*$ the pair dispersion, where $M^*$ is the effective
pair mass.  The latter parameter as well as the pair chemical potential
$\mu_{pair}$ depends on the important, but unknown, gap parameter
$\Delta$ through the fermion self energy $\Sigma$. The decay width
$\Gamma_Q$ is negligibly small for small $Q$ below $T_c$.

While there are alternative ways of deriving the self consistent
equations which we use, (such as a decoupling of the Green's function
equations of motion \cite{Kadanoff}), here we present an approach which
shows how this $GG_0$-based $T$-matrix scheme has strong analogies with
the standard theory of BEC. But, importantly this BEC is embedded in a
self consistent treatment of the fermions.  Physically, one should focus
on $\Delta$ as reflecting the presence of bosonic degrees of freedom. In
the fermionic regime ($\mu>0$), it represents the energy required to
break the pairs, so that $\Delta$ is clearly associated with the
presence of ``bosons". In the bosonic regime, $\Delta^2$ directly
measures the density of pairs.

In analogy with the standard theory of BEC, it is expected \cite{JS3}
that $\Delta$ contains contributions from both noncondensed and
condensed pairs. The associated densities are proportional to
$\Delta^2_{pg}(T)$ and $\tilde{\Delta}_{sc}^2(T)$, respectively.  We may
write the first of several constraints needed to close the set of
equations. (i) One has a constraint on the \textit{total number of
  pairs} \cite{ourreview} which can be viewed as analogous to the usual
BEC number constraint
\begin{equation}
\Delta^2(T) = \tilde{\Delta}_{sc}^2(T) + \Delta_{pg}^2(T) \,.
\label{eq:3}
\end{equation}

To determine $\Delta$, (ii) one imposes the BEC-like constraint that
the pair chemical potential vanishes in the superfluid
phase: 
\begin{equation}
\mu_{pair} =0~~~~T \leq T_c \,.
\label{eq:3d}
\end{equation}
This yields 
\begin{equation}
t_{pg}^{-1}(Q\rightarrow 0) = 0=U_{eff}^{-1}(0) +
\chi(0)
\label{eq:3e}
\end{equation}
so that
\begin{equation}
U_{eff}^{-1}(0) + \sumk \frac{1-2f(\Ek)}{2\Ek} \phik^2 = 0 \,,
\label{eq:Gap}
\end{equation}
Importantly, below $T_c$, $\Delta$ satisfies the usual BCS gap equation.
Here we introduce the quasiparticle dispersion $\Ek = \sqrt{(\ek -
  \mu)^2 + \Delta^2\phik^2}$, 
and $f(x)$ is
the Fermi distribution function.

(iii) In analogy with the standard derivation of BEC,
the total contribution of \textit{noncondensed} pairs is readily computed 
by simply adding up their number, based on
the associated propagator
\begin{equation}
\Delta_{pg}^2 \equiv -  \sum_Q t_{pg}(Q) \,.
\label{eq:PG}
\end{equation}

One can rewrite Eq.~(\ref{eq:PG}) so that it looks more directly like a
number equation, by introducing the Bose distribution function $b(x)$
for noncondensed pairs as
\begin{equation}
\Delta_{pg}^2 = Z^{-1}\sum b(\Omega_q, T) \,,
\label{eq:81}
\end{equation}
so that the noncondensed pair density is given by $Z\Delta_{pg}^2$.
Note that the right hand sides of the previous two equations depend on
the unknown $\Delta$ through the self energy appearing in $G$ which, in
turn enters $t_{pg}$ or $\Omega_q$.  Also note that at $T=0$,
$\Delta_{pg}=0$ so that all pairs are condensed as is consistent with
the mean-field BCS-Leggett ground state.

Finally, in analogy with the standard derivation of BEC, (iv) one can
then compute the number of \textit{condensed} pairs associated with
$\tilde{\Delta}_{sc}$, given that one knows the total $\Delta$ and the
noncondensed component.

Despite this analogy with BEC, fermions are the fundamental particles in
the system. It is their chemical potential $\mu$ that is determined
from the number conservation constraint
\begin{equation}
n = n_f + 2 n_{b0} + 2n_b  \equiv n_f + 2 n_b^{tot}  \,.
\label{eq:Number}
\end{equation}
Here $n_{b0}$ and $n_b$ represent the density of condensed and
noncondensed closed-channel molecules, respectively, $n_b^{tot}$ is the
sum, and $n_f = 2\sum_K G(K)$ is the atomic density associated with the
open-channel fermions.  These closed channel fermions have a
propagator $D(Q)$, which we do not discuss in much detail in
order to make the presentation simpler. 
Here   
\begin{equation}
n_b = -\sum_Q D(Q)  \approx Z_b \sumq b(\Omegaq ) \,,
\label{eq:nb}
\end{equation}
where $b(x)$ is the Bose distribution function.  The renormalized
propagator $D(Q)$ is given by the same equation as
Eq.~(\ref{eq:expandt}) with a different residue $Z^{-1} \rightarrow
Z_b$.

In this way the system of equations is complete \cite{JS3}.  The
numerical scheme is straightforward in principle.  We compute $\Delta$
(and $\mu$) via Eqs.~(\ref{eq:Gap}) and (\ref{eq:Number}), to determine
the contribution from the condensate $\tilde{\Delta}_{sc}$ via
Eqs.~(\ref{eq:3}) and (\ref{eq:PG}).  Above $T_c$, this theory must be
generalized to solve self consistently for $\mu_{pair}$ which no longer
vanishes \cite{ChenThermo}.

\section{Physical Implications: Ultracold Atom Superfluidity}
\label{sec:5}

In this section we compare four distinct classes of experiments on
ultracold trapped fermions with theory. These are thermodynamics
\cite{ThermoScience,ChenThermo}, temperature dependent density profiles
\cite{JS5}, RF pairing gap spectroscopy \cite{Grimm4,Torma1,heyan}, and
collective mode measurements \cite{Thomas2,Grimm3}.  We
address all four experiments in the context of the mean field ground
state of Eq.~(\ref{eq:1a}), and its finite temperature extension
discussed in Section \ref{sec:2B}.  That there appears to be good
agreement between theory and experiment lends rather strong support to
the simple mean field theory, which is at the center of this Review.
Interestingly, pseudogap effects are evident in various ways in these
experiments and this serves to tie the ultracold fermions to the high
$T_c$ superconductors.

\begin{figure}
\includegraphics[width=2.3in,clip]{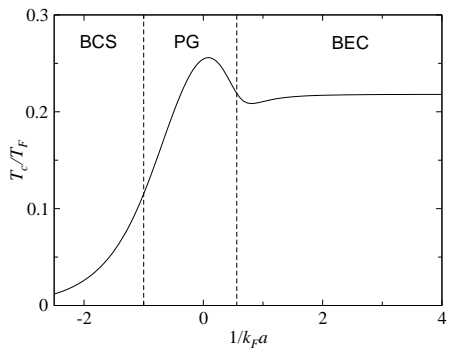}
\caption{Typical behavior of $T_c$ as a function of $1/k_Fa$ in a
  homogeneous system. $T_c$ follows the BCS predictions and approaches
  the BEC asymptote $0.218T_F$ in the BEC limit. In the intermediate
  regime, it reaches a maximum around $1/k_Fa=0$ and a minimum around
  where $\mu=0$.}
\label{fig:Tchomo}
\end{figure}

\subsection{$T_c$ Calculations and Trap Effects}
\label{sec:2E}

\begin{figure}
\includegraphics[width=2.3in,clip]{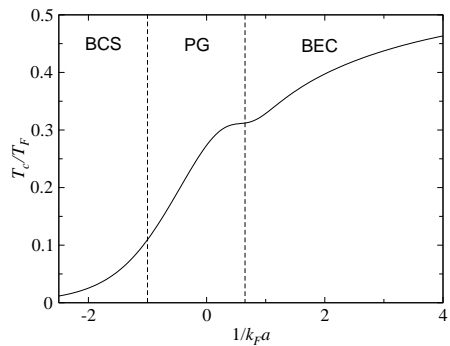}
\caption{Typical behavior of $T_c$ of a Fermi gas in a trap as a
  function of $1/k_Fa$.  It follows BCS prediction in the weak coupling
  limit, $1/k_Fa << -1$, and approaches the BEC asymptote $0.518T_F$ in
  the limit $1/k_Fa \rightarrow +\infty$. In contrast to the homogeneous
  case in Fig.~\ref{fig:Tchomo}, the BEC asymptote is much higher due to a
  compressed profile for trapped bosons.}
\label{fig:Tctrap}
\end{figure}

Before turning to experiment, it is important to discuss the behavior of
the transition temperature which is plotted as a function of scattering
length in Fig.~\ref{fig:Tchomo} for the homogeneous case, presuming
$s$-wave pairing.  We discuss the effects of $d$-wave pairing in Section
\ref{sec:6} in the context of application to the cuprates. Starting from
the BCS regime this figure shows that $T_c$ initially increases as the
interaction strength increases. However, this increase competes with the
opening of a pseudogap or excitation gap $\Delta(T_c)$. Technically, the
pairs become effectively heavier before they form true bound states.
Eventually $T_c$ reaches a maximum (very near unitarity) and then
decreases slightly until field strengths corresponding to the point
where $\mu$ becomes zero.  At this field value (essentially where $T_c$
is minimum), the system becomes a ``bosonic" superfluid, and beyond this
point $T_c$ increases slightly to reach the asymptote corresponding to
an ideal Bose gas.  Trap effects change these results only
quantitatively as seen in Fig.~\ref{fig:Tctrap}.  To treat these trap
effects one introduces the local density approximation (LDA) in which
$T_c$ is computed under the presumption that the chemical potential $\mu
\rightarrow \mu - V(r)$ .  Here we consider a spherical trap with
$V(r)=\frac{1}{2}m\omega^2 r^2$. The Fermi energy $E_F$ is determined by
the total atom number $N$ via $E_F \equiv k_BT_F = \hbar\omega
(3N)^{1/3} \equiv \hbar^2k_F^2/2m$, where $k_F$ is the Fermi wavevector
at the center of the trap. It can be seen that the homogeneous curve is
effectively multiplied by an ``envelope" curve when a trap is present.
This envelope, with a higher BEC asymptote, reflects the fact that the
particle density at the center of the trap is higher in the bosonic,
relative to the fermionic case. In this way $T_c$ is relatively higher
in the BEC regime, as compared to BCS, whenever a trap is present.

\begin{figure}
\includegraphics[width=2.3in,clip]{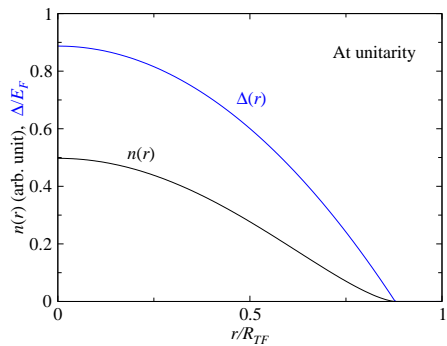}
\caption{Typical spatial profile of $T=0$ density $n(r)$ and fermionic
  excitation gap $\Delta(r)$ of a Fermi gas in a trap. The curves are
  computed at unitarity, where $1/k_Fa=0$. Here $R_{TF}$ is the
  Thomas-Fermi radius.}
\label{fig:denstrap}
\end{figure}

Figure \ref{fig:denstrap} presents a plot of the position dependent
excitation gap $\Delta(r)$ and particle density $n(r)$ profile over the
extent of the trap.  An important point needs to be made: because the
gap is largest at the center of the trap, bosonic excitations will be
dominant there. At the edge of the trap, by contrast, where fermions are
only weakly bound (since $\Delta(r)$ is small), the excitations will be
primarily fermionic.  We will see the implications of these observations
as we examine thermodynamic and RF spectra data in the ultracold gases.

\subsection{Thermodynamical Experiments}

\begin{figure}
\includegraphics[width=2.8in,clip]{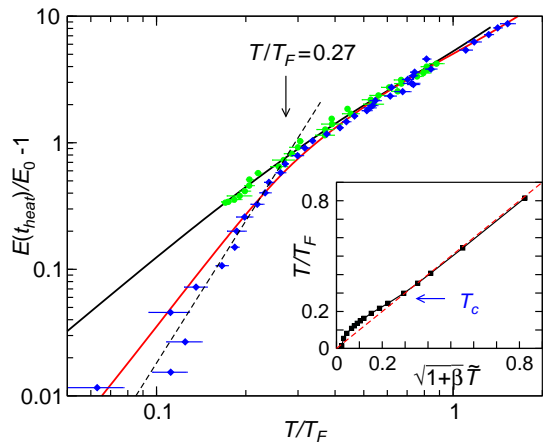}
\caption{(color) Energy $E$ vs physical temperature $T$. The
upper curve and data points correspond to the BCS or essentially
free Fermi gas case, and the lower curve and data correspond to
unitarity. The latter provide indications for a phase transition.
The inset shows how temperature must be recalibrated below
$T_c$. From Ref.~\cite{ThermoScience}.}
\label{fig:231new}
\end{figure}

Figure \ref{fig:231new} present a plot which compares experiment and
theory in the context of thermodynamic experiments
\cite{ThermoScience,ChenThermo} on trapped fermions. Plotted on the
vertical axis is the energy which can be input in a controlled fashion
experimentally. The horizontal axis is temperature which is calibrated
theoretically based on an effective temperature $\tilde{T}$ introduced
phenomenologically, and discussed below. The experimental data are shown
for the (effectively) non-interacting case as well as unitary.  In this
discussion we treat the non-interacting and BCS cases as essentially
equivalent since $\Delta$ is so small on the scale of the temperatures
considered.  The solid curves correspond to theory for the two cases.
Although not shown here, even without a temperature calibration, the
data suggests a phase transition is present in the unitary case. This
can be seen as a result of the change in slope of $E(\tilde{T})$ as a
function of $\tilde{T}$.

The phenomenological temperature $\tilde{T}$ is relatively easy to
understand.  What was done experimentally to deduce this temperature was
to treat the unitary case as an essentially free Fermi gas to, thereby,
infer the temperature from the width of the density profiles, but with
one important proviso: a numerical constant is introduced to account for
the fact that the density profiles become progressively narrower as the
system varies from BCS to BEC.  This systematic variation in the profile
widths reflects the fact that in the free Fermi gas case, Pauli
principle repulsion leads to a larger spread in the particle density
than in the bosonic case.  And the unitary regime has a profile width
which is somewhere in between, so that one parametrizes this width by a
simple function of $\beta$.  We can think of $\beta$ as reflecting
bosonic degrees of freedom, within an otherwise fermionic system. At
$\beta \equiv 0$ the system is a free Fermi gas.  The principle
underlying this rescaling of the non-interacting gas is known as the
``universality hypothesis" \cite{JasonHo,Thomasnew}.  At unitarity, the
Fermi energy of the non-interacting system is the only energy scale in
the problem (for the widely used contact potential) since all other
scales associated with the two-body potential drop out when $ a_s
\rightarrow \pm \infty $.
We refer to this phenomenological fitting temperature procedure as
Thomas-Fermi (TF) fits.

An interesting challenge was to relate this phenomenological temperature
$\tilde{T}$ to the physical temperature $T$; more precisely one compares
$\sqrt{1 + \beta} \tilde{T}$ and $T$.  This relationship is demonstrated
in the inset of Fig.~\ref{fig:231new}.  And it was in this way that
the theory and experiment could be plotted on the same figure, as shown
in the main body of Fig.~\ref{fig:231new}. The inset was obtained using
theory only. The theoretically produced profiles were analyzed just as
the experimental ones to extract $\sqrt{1 + \beta} \tilde{T}$ and
compare it to the actual $T$. Above $T_c$ no recalibration was needed as
shown by the straight line going through the diagonal.  Below $T_c$ the
phenomenologically deduced temperatures were consistently lower than the
physical temperature.  That the normal state temperatures needed no
adjustment shows that the phenomenology captures important physics.  It
misses, however, an effect associated with the presence of a condensate
which we will discuss shortly.

\begin{figure}
\includegraphics[width=2.8in,clip]{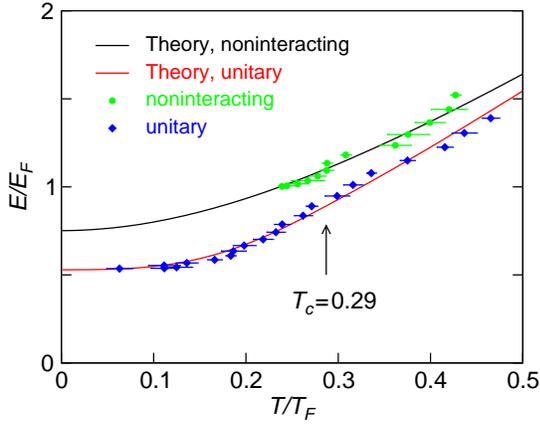}
\caption{(color) Low temperature comparison of theory (curves) and experiments
  (symbols) in terms of $E/E_F$ ($E_F=k_BT_F$) per atom as a function of
  $T/T_F$, for both unitary and noninteracting gases in a Gaussian trap.
  From Ref.~\onlinecite{ThermoScience}.}
\label{fig:230new}
\end{figure}

We next turn to a more detailed comparison of theory and experiment for
the global and low $T$ thermodynamics.  Figure \ref{fig:230new} presents
a blow-up of $E$ at the lowest $T$ comparing the unitary and
non-interacting regimes.  The agreement between theory and experiment is
quite good.  In the figure, the temperature dependence of $E$ reflects
primarily fermionic excitations at the edge of the trap, although there
is a small bosonic contribution as well.  It should be noted that the
theoretical plots were based on fitting $\beta$ to experiment by picking
a magnetic field very slightly off resonance. [In the simple mean field
theory $\beta = -0.41$, and in Monte Carlo simulations \cite{Carlson3}
$\beta = -0.54$. Both these theoretical numbers lie on either side of
experiment \cite{ThermoScience} where $\beta = -0.49$].

\begin{figure}
\includegraphics[width=2.8in,clip]{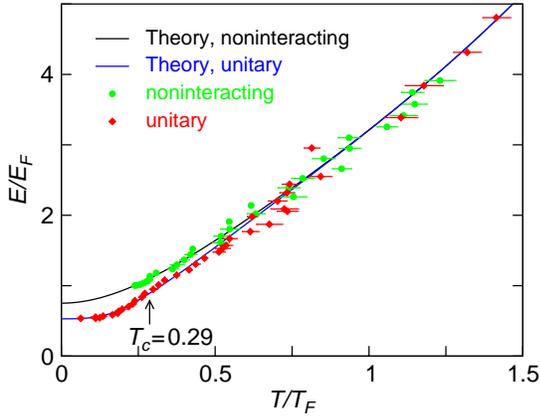}
\caption{(color) Same as Fig.~\ref{fig:230new} but for a much larger range of
  temperature. The quantitative agreement between theory and experiment
  is very good.The fact that the two experimental (and the two
    theoretical) curves do not merge until higher $T^* > T_c$ is
      consistent with the presence of a pseudogap.}
\label{fig:232new}
\end{figure}

Figure \ref{fig:232new} presents a wider temperature scale plot which,
again, shows very good agreement. Importantly one can see the effect of
a pseudogap in the unitary case. The temperature $T^*$ can be picked out
from the plots as that at which the non-interacting and unitary curves
intersect.  This corresponds roughly to $T^* \approx 2 T_c$.

\subsection{Temperature Dependent Particle Density Profiles}

\begin{figure}
\includegraphics[width=2.8in,clip]{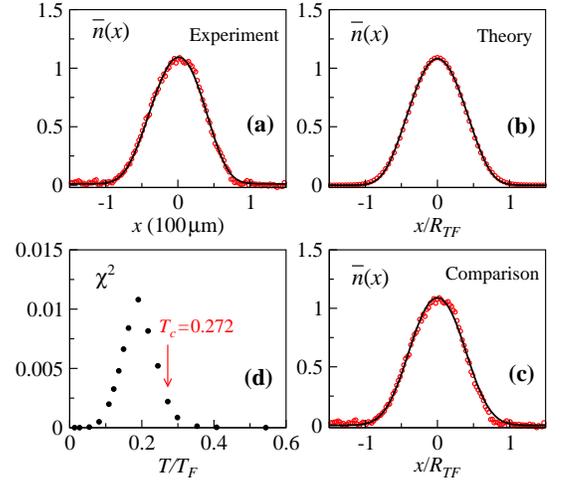}
\caption{(color) Temperature dependence of (a) experimental one-dimensional
  spatial profiles (circles) and TF fit (line) from
  Ref.~\onlinecite{Kinast}, (b) TF fits (line) to theory both at $T
  \approx 0.7T_c \approx 0.19T_F$ (circles) and (c) overlay of
  experimental (circles) and theoretical (line) profiles, as well as (d)
  relative rms deviations ($\chi^2$) associated with these fits to
  theory at unitarity.  The circles in (b) are shown as the line in (c).
  The profiles have been normalized so that $N=\int \bar{n}(x) dx = 1$,
  and we set $R_{TF} = 100~\mu$m in order to overlay the two curves.
  $\chi^2$ reaches a maximum around $T=0.19T_F$.}
\label{fig:4new}
\end{figure}

In order to understand more deeply the behavior of the thermodynamics,
we turn next to a comparison of finite $T$ density profiles. Experiments
which measure these profiles \cite{Grimm2,Kinast} all report that they
are quite smooth at unitarity, without any signs of the bimodality seen
in the BEC regime. We discuss these profiles in terms of the four panels
in Fig.~\ref{fig:4new}. These figures are a first step in understanding
the previous temperature calibration procedure.

In this figure we compare theory and experiment for the unitary case.
The experimental data were estimated to correspond to roughly this same
temperature ($T/T_F = 0.19$) based on the calibration procedure
discussed above.  The profiles shown are well within the superfluid
phase ($T_c \approx 0.3T_F$ at unitarity).  This figure presents
Thomas-Fermi fits \cite{Kinast} to (a) the experimental and (b)
theoretical profiles as well as (c) their comparison, for a chosen
$R_{TF} = 100~\mu$m, which makes it possible to overlay the
experimental data (circles) and theoretical curve (line).  Finally
Fig.~\ref{fig:4new}d indicates the relative $\chi^2$ or root-mean-square
(rms) deviations for these TF fits to theory.  This figure was made in
collaboration with the authors of Ref.~\cite{Kinast}.  Two of the three
dimensions of the theoretical trap profiles were integrated out to
obtain a one-dimensional representation of the density distribution
along the transverse direction: $\bar {n} (x) \equiv \int dydz \, n(r)$.

This figure is in contrast to earlier theoretical studies which predict
a significant kink at the condensate edge which appears not to have been
seen experimentally \cite{Grimm2,Kinast}.  Moreover, the curves behave
monotonically with both temperature and radius.  Indeed, in the unitary
regime the generalized TF fitting procedure of Ref.~\onlinecite{Kinast}
works surprisingly well.  And these reasonable TF fits apply to
essentially all temperatures investigated experimentally \cite{Kinast},
as well as theoretically, including in the normal state.

It is important to establish why the profiles are so smooth, and the
condensate is, in some sense, rather invisible, except for its effect on
the TF-inferred-temperature.  This apparent smoothness can be traced to
the presence of noncondensed pairs of fermions which need to be included
in any consistent treatment. Indeed, these pairs below $T_c$ are a
natural counterpart of the pairs above $T_c$ which give rise to
pseudogap effects.

\begin{figure}
\includegraphics[width=3.2in,clip]{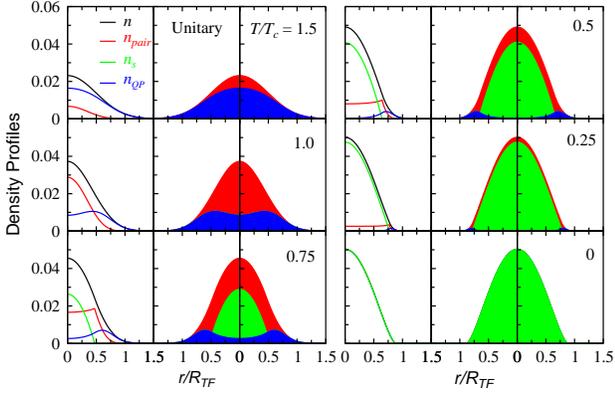}
\caption{(color) Decomposition of density profiles at various
  temperatures at unitarity. Here green (light gray) refers to the
  condensate, red (dark gray) to the noncondensed pairs and blue (black)
  to the excited fermionic states. $T_c = 0.27T_F$, and $R_{TF}$ is the
  Thomas-Fermi radius.}
\label{fig:23new}
\end{figure}

To see how the various contributions enter into the trap profile, in
Fig.~\ref{fig:23new} we plot a decomposition of this profile for
various temperatures from below to above $T_c$.  The various color codes
indicate the condensate along with the noncondensed pairs and the
fermions. This decomposition is based on the superfluid density so that
all atoms participate in the condensation at $T=0$. This, then, forms
the basis for addressing both thermodynamics and RF pairing-gap
spectroscopy in this Review.

The figure shows that by $T = T_c/2$ there is a reasonable number of
excited fermions and bosons.  As anticipated earlier in Section
\ref{sec:2B}, the latter are at the trap edge and the former in the
center. By $ T= T_c$ the condensate has disappeared and the excitations
are a mix of fermions (at the edge) and bosons towards the center.
Indeed, the noncondensed bosons are still present by $ T = 1.5 T_c$, as
a manifestation of a pseudogap effect.  Only for somewhat higher $T
\approx 2 T_c$ do they disappear altogether, so that the system becomes
a non-interacting Fermi gas.

Two important points should be made.  The noncondensed pairs clearly are
responsible for smoothing out what otherwise would be a discontinuity
\cite{Chiofalo,JasonHo} between the fermionic and condensate
contributions.  Moreover, the condensate shrinks to the center of the
trap as $T$ is progressively raised. It is this thermal effect which is
responsible for the fact that the TF fitting procedure for extracting
temperature leads to an underestimate as shown in the inset to
Fig.~\ref{fig:231new}.  The presence of the condensate tends to make the
atomic cloud smaller so that the temperature appears to be lower in the
TF fits.

\subsection{RF Pairing gap Spectroscopy}

\begin{figure}
\includegraphics[width=2.6in,clip]{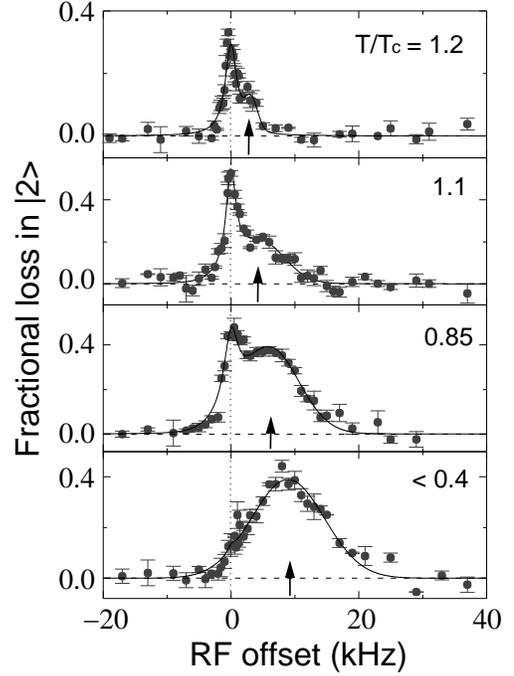}
\caption{Experimental RF Spectra at unitarity. The temperatures labeled
  in the figure were computed theoretically at unitarity based on
  adiabatic sweeps from BEC.  The two top curves, thus, correspond to
  the normal phase, thereby, indicating pseudogap effects. Here $E_F =
  2.5 \mu$K, or 52 kHz.  From Ref.~\onlinecite{Grimm4}.}
\label{fig:24Cheng}
\end{figure}

Measurements \cite{Grimm4} of the excitation gap $\Delta$ have been
made by using a third atomic level, called $|3 \rangle$, which does not
participate in the superfluid pairing. Under application of RF fields,
one component of the Cooper pairs, called $|2 \rangle$, is presumably
excited to state $|3\rangle$.  If there is no gap $\Delta$ then the
energy it takes to excite $|2 \rangle$ to $|3 \rangle$ is the atomic
level splitting $\omega_{23}$. In the presence of pairing (either above
or below $T_c$) an extra energy $\Delta$ must be input to excite the
state $|2 \rangle$, as a result of the breaking of the pairs.  Figure
\ref{fig:24Cheng} shows a plot of the spectra near unitarity for four
different temperatures, which we discuss in more detail below.  In
general for this case, as well as for the BCS and BEC limits, there are
two peak structures which appear in the data: the sharp peak
at $\omega_{23} \equiv 0$ which is associated with ``free" fermions at
the trap edge and the broader peak which reflects the presence of
paired atoms; more directly this broad peak derives from the distribution of
$\Delta$ in the trap.  At high $T$ (compared to $\Delta$), only the
sharp feature is present, whereas at low $T$ only the broad feature
remains.  The sharpness of the free atom peak can be understood as
coming from a large phase space contribution associated with the $2
\rightarrow 3$ excitations \cite{heyan}.  Clearly, these data alone do not
directly indicate the presence of superfluidity, but rather they provide
strong evidence for pairing.

\begin{figure}
\centerline{\includegraphics[clip,width=2.5in,height=2.8in]{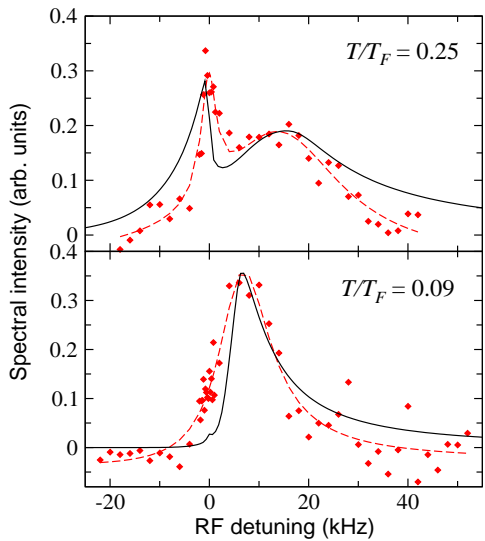}}
\caption{Comparison of calculated RF spectra (solid curve, $T_c \approx
  0.29T_F$) with experiment (symbols) in a harmonic trap calculated at
  822 G for the two lower temperatures. The temperatures were chosen
  based on Ref.~\cite{Grimm4}. The particle number was reduced by a
  factor of 2, as found to be necessary in addressing another class of
  experiments \cite{Strinati5}.  The dashed lines are a guide to the
  eye. From Ref.~\onlinecite{heyan}.}
\label{fig:100}
\end{figure}

As pointed out in Ref. \cite{Torma1} these experiments serve as a
counterpart to superconducting tunneling in providing information about
the excitation gap. A theoretical understanding of these data was first
presented in Ref.~\onlinecite{Torma2} using the framework of Section
\ref{sec:2B}.  Subsequent work \cite{heyan} addressed these data in a
more quantitative fashion as plotted in Fig.~\ref{fig:100}.  Here the
upper and lower panels correspond respectively to intermediate and low
temperatures. For the latter one sees that the sharp ``free atom" peak
has disappeared, so that fermions at the edge of the trap are
effectively bound at these low $T$.  Agreement between theory and
experiment is quite satisfactory, although the total number of particles
was adjusted somewhat relative to the experimental estimates.

It is interesting to return to the previous figure
(Fig.~\ref{fig:24Cheng}) and to discuss the temperatures in the various
panels. What is measured experimentally are temperatures $T'$ which
correspond to the temperature at the start of a sweep from the BEC limit
to unitarity. Here fits to the BEC-like profiles are used to deduce $T'$
from the shape of the Gaussian tails in the trap. Based on knowledge
about thermodynamics (entropy $S$), and given $T'$, one can then compute
the final temperature in the unitary regime, assuming $S$ is constant.
Indeed, this adiabaticity has been confirmed experimentally in related
work \cite{Grimm2}. We find that the four temperatures are as indicated
in the figures.  Importantly, one can conclude that the first two cases
correspond to a normal state, albeit close to $T_c$.  Importantly, these
figures suggest that a pseudogap is present as reflected by the broad
shoulder above the narrow free atom peak.

\subsection{Collective Breathing Modes at $T \approx 0$}

\begin{figure}
\includegraphics[angle=270,width=2.8in,clip]{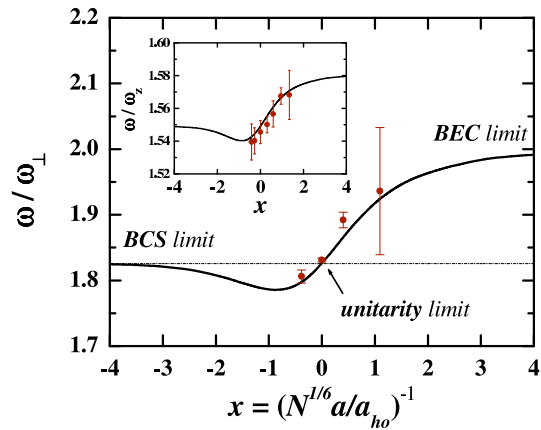}
\caption{Breathing mode frequencies as a function of $\kappa \approx
  1.695 (k_Fa)^{-1}$, from Tosi \textit{et al.} \cite{Tosi}.  The main
  figure  and inset plot the transverse and axial frequencies,
  respectively. The   solid curves are calculations \cite{Tosi} based on
  BCS-BEC crossover   theory at $T=0$, and the symbols plot the
  experimental data from Kinast   \textit{et al.}  \cite{Thomas2}. 
}
\label{fig:24Tosi}
\end{figure}

We turn, finally, to a comparison between theory \cite{Tosi,Heiselberg}
and experiment \cite{Thomas2,Grimm3,Kinast2,Thomasnew} for the
collective breathing modes within a trap at $T\approx 0$. The very good
agreement has provided some of the earliest and strongest support for
the simple mean field theory of Eq.~(\ref{eq:1a}).  Interestingly, Monte
Carlo simulations which initially were viewed as a superior approach,
lead to significant disagreement between theory and experiment
\cite{Grimmprivate}.  Shown in Fig.~\ref{fig:24Tosi} is this comparison
for the axial mode in the inset and the radial mode in the main body of
the figure as a function of magnetic field.  The experimental data are
from Ref.~\onlinecite{Thomas2}. The original data on the radial modes from
Ref.~\onlinecite{Grimm3}, was in disagreement with that
of Ref.~\onlinecite{Thomas2}, but this has since been
corrected \cite{Grimmprivate}, and there is now
a consistent experimental picture from both the Duke and the Innsbruck
groups for the radial mode frequencies.

At $T=0$, calculations of the mode frequencies can be reduced to a
calculation of an equation of state for $\mu$ as a function of $n$.  One
of the most important conclusions from this figure is that the behavior
in the near-BEC limit (which is still far from the BEC asymptote) shows
that the mode frequencies \textit{decrease} with increasing magnetic
field. This is opposite to earlier predictions \cite{Stringari2} based
on the behavior of true bosons where a Lee-Yang term would lead to an
increase. Indeed, the pair operators do not obey the commutation
relations of true bosons except in the zero density or $k_Fa \rightarrow
0^+$ limit \cite{Schrieffer}. Figure \ref{fig:24Tosi}, thus, underlines
the fact that fermionic degrees of freedom (or compositeness) are still
playing a role at these magnetic fields.  There are predictions in the
literature \cite{Tan2} that one needs to achieve $k_Fa$ somewhat less
than $0.3$ (experimentally, the smallest values for these experiments
are 0.3 and 0.7 for the various groups) in order to approach the true
bosonic limit.  At this point, then, the simple mean field theory will
no longer be adequate.  Indeed, there are other indications
\cite{Petrov} of the breakdown of this mean field in the extreme BEC
limit which are, physically, reflected in the width of the particle
density profiles.  This originates from an overestimate (by roughly a
factor of 3) of the size of the effective ``inter-boson" scattering
length.

Overall the mean field theory presented here looks very promising.
Indeed, the agreement between theory and experiment is better than one
might have anticipated.  For the collective mode frequencies, it appears
to be better than Monte Carlo calculations \cite{Grimmprivate}.
Nevertheless, uncertainties remain.  Theories which posit a different
ground state will need to be compared with the four experiments
discussed here. It is, finally, quite possible that incomplete $T=0$
condensation will become evident in future experiments.  If so, an
alternative wavefunction will have to be contemplated
\cite{Tan2,Holland2}.  What appears to be clear from the current
experiments is that, just as in high $T_c$ superconductors, the
ultracold fermionic superfluids exhibit pseudogap effects. These are
seen in thermodynamics, in RF spectra and in the temperature dependence
of the profiles (through the noncondensed pair contributions).
Moreover, while not discussed here, at finite $T$, damping of the
collective mode frequencies seems to change qualitatively
\cite{Thomasnew} at a temperature which is close to the estimated $T^*$.

Looking to the future, at an experimental level, new pairing gap
spectroscopies appear to be emerging at a fairly rapid pace
\cite{Jin5,Hulet4}.  These will further test the present and subsequent
theories. Indeed, recently, a probe of the closed channel fraction
\cite{Hulet4} has been analyzed \cite{ChenClosed} within the present
framework and has led to good quantitative agreement between theory and
experiment.

\section{Physical Implications: High $T_c$ Superconductivity} 
\label{sec:6}

\subsection{Phase Diagram and Superconducting Coherence}
\label{sec:6A}

\begin{figure}
\includegraphics[width=2.8in,clip]{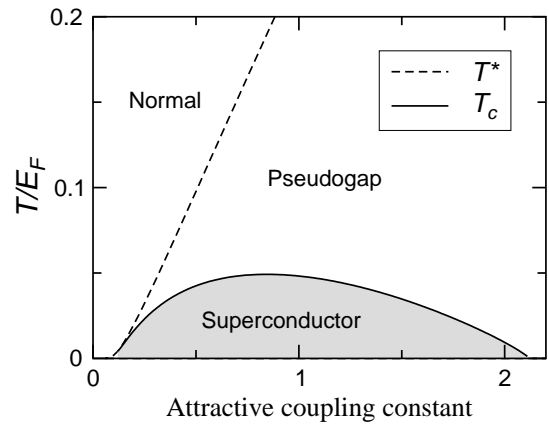}
\caption{Typical phase diagram for a quasi-two dimensional $d$-wave
  superconductor on a tight-binding lattice at high filling $n\approx
  0.85$ per unit cell; here the horizontal axis corresponds to
  $-U/4t$, where $t$ is the in-plane hopping matrix element.}
\label{fig:23}
\end{figure}

The high $T_c$ superconductors are different from the ultracold
fermionic superfluids in one key respect; they are $d$-wave
superconductors and their electronic dispersion is associated with a
quasi-two dimensional tight binding lattice. In many ways this is not a
profound difference from the perspective of BCS-BEC crossover.  Figure
\ref{fig:23} shows a plot of the two important temperatures $T_c$ and
$T^*$ as a function of increasing attractive coupling.  On the left is
BCS and the right is PG. The BEC regime is not visible. This is because
$T_c$ disappears before it can be accessed.  This disappearance of $T_c$
is relatively easy to understand.  Because the $d$-wave pairs are more
extended (than their $s$-wave counterparts) they experience Pauli
principle repulsion more intensely. Consequently the pairs localize
(their mass is infinite) well before the fermionic chemical potential is
negative \cite{Chen1}.

The competition between $T^*$ and $T_c$, in which as $T^*$ increases,
$T_c$ decreases, is also apparent in Fig.~\ref{fig:23}. This is a
consequence of pseudogap effects.  More specifically, the pairs become
heavier as the gap increases in the fermionic spectrum, competing with
the increase of $T_c$ due to the increasing pairing strength.  It is
interesting to compare Fig.~\ref{fig:23} with the experimental phase
diagram plotted as a function of the doping concentration $x$ in
Fig.~\ref{fig:5}.  If one inverts the horizontal axis (and ignores the
unimportant AFM region) the two are very similar. To make an association
from coupling $U$ to the variable $x$, it is reasonable to fit $T^*$. It
is not particularly useful to implement this last step here, since we
wish to emphasize crossover effects which are not complicated by ``Mott
physics".

\begin{figure*}
\centerline{\includegraphics{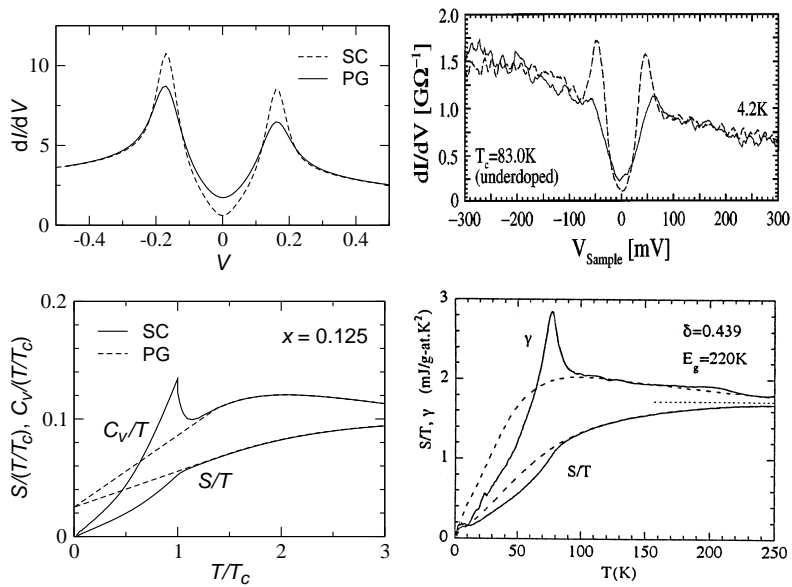}}
\caption{Extrapolated normal state (PG) and superconducting state (SC)
  contributions to SIN tunneling and thermodynamics (left), as well as
  comparison with experiments (right) on tunneling for BSCCO
  \cite{Renner} and on specific heat for
  Y$_{0.8}$Ca$_{0.2}$Ba$_2$Cu$_3$O$_{7-\delta}$ \cite{Loram98}.  The
  theoretical SIN curve is calculated for $T = T_c/2$, while the
  experimental curves are measured outside (dashed line) and inside
  (solid line) a vortex core.  }
\label{fig:25}
\end{figure*}

Because of quasi-two dimensionality, the energy scales of the vertical
axis in Fig.~\ref{fig:23} are considerably smaller than their three
dimensional analogues.  Thus, pseudogap effects are intensified, just as
conventional fluctuation effects are more apparent in low dimensional
systems.  This may be one of the reasons why the cuprates are among the
first materials to clearly reveal pseudogap physics.  Moreover, the
present calculations show that in a strictly 2D material, $T_c$ is
driven to zero, by bosonic or fluctuation effects.  This is a direct
reflection of the fact that there is no Bose condensation in 2D.

The presence of pseudogap effects raises an interesting set of issues
surrounding the signatures of the transition which the high $T_c$
community has wrestled with, much as the cold atom community is doing
today.  For a charged superconductor there is no difficulty in measuring
the superfluid density, through the electrodynamic response. Thus one
knows with certainty where $T_c$ is.  Nevertheless, people have been
concerned about precisely how the onset of phase coherence is reflected
in thermodynamics, such as $C_V$ or in the fermionic spectral function,
given that a gap is already present at the onset of superconductivity.
One understands how phase coherence shows up in BCS theory, since the
ordered state is always accompanied by the appearance of an excitation
gap. 

To address these coherence effects one has to introduce a distinction
\cite{Chen4} between the self energy associated with noncondensed and
condensed pairs. This distinction is blurred by the approximations made
in Section \ref{sec:2B}.  Within this improved scheme \cite{Chen4}
superconducting coherence effects can be probed as, presented in
Fig.~\ref{fig:25}, along with a comparison to experiment.  Shown are the
results of specific heat and tunneling calculations and their
experimental counterparts \cite{LoramPhysicaC,Renner}.  The latter
measures, effectively, the density of fermionic states.  Here the label
``PG" corresponds to an extrapolated normal state in which we set the
superconducting order parameter $\Delta_{sc}$ to zero, but maintain the
the total excitation gap $\Delta$ to be the same as in a phase coherent,
superconducting state.  Agreement between theory and experiment is
satisfactory.

\subsection{Electrodynamics in the superconducting phase}

In some ways the subtleties of phase coherent pairing are even more
perplexing in the context of electrodynamics.  Figure \ref{fig:10}
presents a paradox in which the excitation gap for fermions appears to
have little to do with the behavior of the superfluid density.  This
superfluid density can be readily computed within the BCS-BEC crossover
scenario \cite{Chen2,JS1}.  Particularly important is to include all
excitations of the condensate in a fully consistent fashion, compatible
with thermodynamics, and which is also manifestly gauge invariant.  To
make contact with electrodynamic experiments, one has to introduce the
variable $x$ and this is done via a fit to $T^*(x)$ in the phase
diagram.  In addition it is also necessary to fit $\rho_s(T=0,x)$ to
experiment, and we do so here, noting that \cite{Uemura} the Uemura
relation $\rho_s(0,x) \propto T_c(x)$ no longer holds for very
underdoped samples \cite{Hardy3,Lemberger2}.  By fitting these
$x$-dependent quantities we are, in effect accounting for at least some
aspects of Mott physics.  The paradox raised by Fig.~\ref{fig:10} is
resolved by noting that there are bosonic excitations of the condensate
\cite{Chen2} and that these become more marked with underdoping, as
pseudogap effects increase.  In this way $\rho_s$ does not exclusively
reflect the fermionic gap, but rather vanishes ``prematurely" before
this gap is zero, as a result of pair excitations of the condensate.

\begin{figure*}
\centerline{\includegraphics{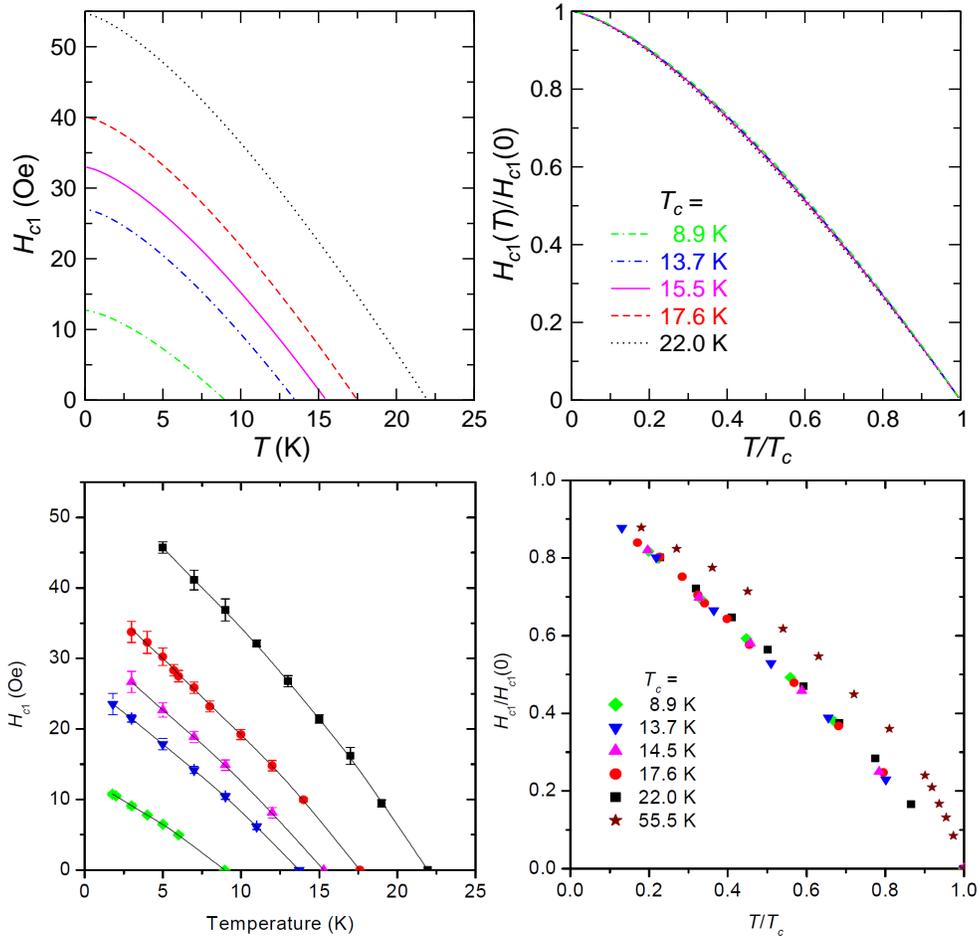}}
\caption{(color) Comparison between calculated lower critical field, $H_{c1}$,
  as a function of $T$ (upper left panel), and experimental data (lower
  left) from Ref.~\onlinecite{Hardy3}, with variable doping concentration
  $x$.  The right column shows normalized plots, $H_{c1}(T)/H_{c1}(0)$
  versus $T/T_c$, for theory and experiment, respectively, revealing a
  quasi-universal behavior with respect to doping, with the exception of
  the $T_c = 55.5K$ ortho-II phase. Both theory plots share the same
  legends. The quantitative agreement between theory and experiment is
  quite good.  }
\label{fig:26}
\end{figure*}

This theory can be quantitatively compared with experiment. Figure
\ref{fig:26} presents theoretical and experimental plots of the lower
critical field, $H_{c1}(T)$, for a group of severely underdoped YBCO
crystals as considered in Ref.~\onlinecite{Hardy3}. There it was argued
that $H_{c1}(T) \propto \rho_s(T)$, so that the lower critical field
effectively measures the in-plane superfluid density.  Experimentally
what is directly measured is the magnetization with applied field
parallel to the $c$-axis.  The experimental results are shown on the
lower two panels and theory on the upper two.  The left hand figures
plot $H_{c1}(T)$ vs $T$ and the right hand figures correspond to a
rescaling of this function in the form $H_{c1}(T)/H_{c1}(0)$ vs $T/T_c$.
Theoretically, it is found that the fermionic contribution leads to a
linear $T$ dependence at low $T$, associated with $d$-wave pairing,
whereas the bosonic term introduces a $T^{3/2}$ term.  Quite remarkably
even when the Uemura relation no longer holds, there is still a
``universality" in the normalized plots as shown in both theory and
experiment by the right hand figures. It should be noted that the
experimental plot contains (at $T_c = 55.5K$) a slightly different
cuprate phase known as the ortho-II phase, which does not lie on the
universal curves.  The universality found here can be understood as
associated with the fact that $T_c(x)$, rather than $\Delta(x)$, is the
fundamental energy scale in $\rho_s(T,x)$.  The reason that $\Delta(x)$
is not the sole energy scale is that bosonic degrees of freedom are also
present, and help to drive $\rho_s$ to zero at $T_c$.  By contrast,
Fermi-liquid based approaches \cite{Lee,Hardy3} assume that the fermions
are the only relevant excitations, and they account for this data by
introducing a phenomenological parameter $\alpha$ which corresponds to
the effective charge of the fermionic quasi-particles.

\begin{figure}
\centerline{\includegraphics[width=3.05in,clip]{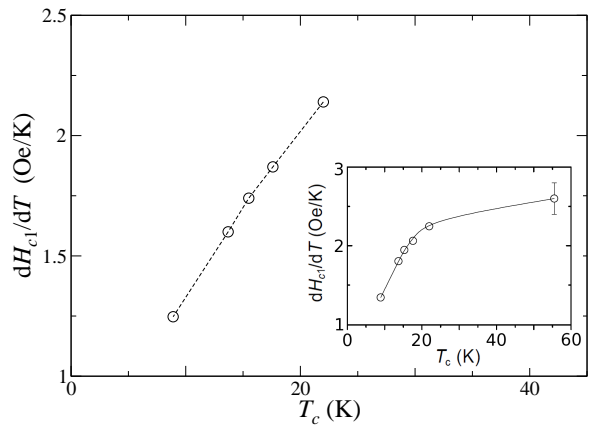}}
\caption{Comparison of theoretically calculated low $T$ slope
  $\mathrm{d} H_{c1}/\mathrm{d} T$ (main figure) for various doping
  concentrations (corresponding to different $T_c$) in the underdoped
  regime with experimental data (inset) from Ref.~\onlinecite{Hardy3}. The
  theoretical slopes are estimated using the low temperature data points
  accessed experimentally. The quantitative agreement is very good.  }
\label{fig:27}
\end{figure}

As anticipated in earlier theoretical calculations \cite{Chen2} the
bosonic contribution begins to dominate in severely underdoped systems
so that the slope $dH_{c1}/dT $ (associated with the lowest temperatures
reached experimentally) should \textit{decrease} with underdoping.
Although observed a number of years after this prediction, this is
precisely what is seen experimentally, as shown in Fig.~\ref{fig:27}.
Here the inset plots the experimental counterpart data.  It can be seen
that theory and experiment are in reasonably good quantitative
agreement.  This theoretical viewpoint is very different from a
``Fermi-liquid" based treatment of the superconducting state, for which
the strong decrease in the slope of $H_{c1}$ was not expected.
Within the present formalism, the optical conductivity \cite{Timusk}
$\sigma(\omega)$ is similarly modified \cite{Iyengar} to include bosonic
as well as fermionic contributions.

\subsection{Bosonic Power Laws and Pairbreaking Effects}
\label{sec:6B}

The existence of noncondensed pair states below $T_c$ affects
thermodynamics, in the same way that electrodynamics is affected, as
discussed above. Moreover, one can predict \cite{Chen3} that the $q^2$
dispersion will lead to ideal Bose gas power laws in thermodynamical and
transport properties.  These will be present in addition to the usual
power laws or (for $s$-wave) exponential temperature dependencies
associated with the fermionic quasi-particles.  Note that the $q^2$
dependence is dictated by the ground state of Eq.~(\ref{eq:1a}).  Clearly
this mean field like state is inapplicable in the extreme BEC limit,
where, presumably inter-boson effects become important and lead to a
linear dispersion.  Presumably, in the PG or near-BEC regimes, fermionic
degrees of freedom are still dominant and it is reasonable to apply
Eq.~(\ref{eq:1a}).  Importantly, at present neither the cuprates nor the
cold atom systems access this true BEC regime.

The consequences of these observations can now be listed \cite{Chen3}.
For a quasi-two dimensional system, $C_v/T$ will appear roughly constant
at the lowest temperatures, although it vanishes strictly at $T=0$ as
$T^{1/2}$. The superfluid density $\rho_s(T)$ will acquire a $T^{3/2}$
contribution in addition to the usual fermionic terms. By contrast, for
spin singlet states, there is no explicit pair contribution to the
Knight shift. In this way the low $T$ Knight shift reflects only the
fermions and exhibits a scaling with $T/\Delta(0)$ at low temperatures.
Experimentally, in the cuprates, one usually sees a finite low $T$
contribution to $C_v/T$.  A Knight shift scaling is seen.  Finally, also
observed is a deviation from the predicted $d$-wave linear in $T$ power
law in $\rho_s$.  The new power laws in $C_v$ and $\rho_s$ are
conventionally attributed to impurity effects. Experiments are not yet
at a stage to clearly distinguish between these two alternative
explanations.

Pairbreaking effects are viewed as providing important insight into the
origin of the cuprate pseudogap.  Indeed, the different pairbreaking
sensitivities of $T^*$ and $T_c$ are usually proposed to support the
notion that the pseudogap has nothing to do with superconductivity.  To
counter this incorrect inference, a detailed set of studies was
conducted, (based on the BEC-BCS scenario), of pairbreaking in the
presence of impurities \cite{Chen-Schrieffer,Kao4} and of magnetic
fields \cite{Kao3}. These studies make it clear that the superconducting
coherence temperature $T_c$ is far more sensitive to pairbreaking than
is the pseudogap onset temperature $T^*$. Indeed, the phase diagram of
Fig.~\ref{fig:23} which mirrors its experimental counterpart, shows the
very different, even competing nature of $T^*$ and $T_c$, despite the
fact that both arise from the same pairing correlations.

\subsection{Anomalous Normal State Transport: Nernst
Coefficient}
\label{sec:6c}

\begin{figure}
\includegraphics[width=3in]{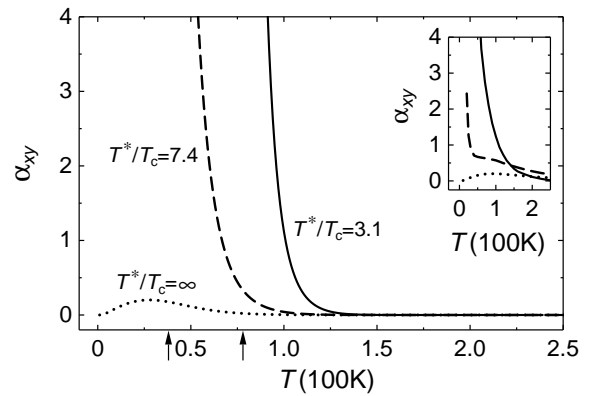}
\caption{Calculated transverse thermoelectric response, which appears in
  the Nernst coefficient, as a function of temperature for the underdoped
  cuprates.}
\label{fig:33}
\end{figure}

Much attention is given to the anomalous behavior of the Nernst
coefficient in the cuprates \cite{Nernst}.  This coefficient is rather
simply related to the transverse thermoelectric coefficient
$\alpha_{xy}$ which is plotted in Fig.~\ref{fig:33}.  In large part, the
origin of the excitement in the literature stems from the fact that the
Nernst coefficient behaves smoothly through the superconducting
transition. Below $T_c$ it is understood to be associated with
superconducting vortices. Above $T_c$ if the system were a Fermi liquid,
there are arguments to prove that the Nernst coefficient should be
essentially zero.  Hence the observation of a non-negligible Nernst
contribution has led to the picture of fluctuating ``normal state
vortices".

The formalism of Ref.~\onlinecite{Tan} can be used to address these data
within the framework of BCS-BEC crossover.  The results are plotted in
Fig.~\ref{fig:33} with a subset of the data plotted in the upper right
inset.  It can be seen that the agreement is reasonable. In this way a
``pre-formed pair" picture appears to be a viable alternative to
``normal state vortices".  It will, ultimately, be necessary to take
these transport calculations below $T_c$.  This is a project for future
research and in this context it will be important to establish in this
picture how superconducting state vortices are affected by the
noncondensed pairs and conversely.

\section{Conclusions}

In this Review we have summarized a large body of work on the subject of
the BCS-BEC crossover scenario. In this context, we explored the
intersection of two very different fields: high $T_c$ superconductivity
and cold atom superfluidity.  Theories of cuprate superconductivity can
be crudely classified as focusing on ``Mott physics" which reflects the
anomalously small zero temperature superfluid density and ``crossover
physics", which reflects the anomalously short coherence length. Both
schools are currently very interested in explaining the origin of the
mysterious pseudogap phase. In this Review we have presented a case for
its origin in crossover physics.  The pseudogap in the normal state can
be associated with meta-stable pairs of fermions; a (pseudogap) energy
must be supplied to break these pairs apart into their separate
components.  The pseudogap also persists below $T_c$ in the sense that
there are noncondensed fermion pair excitations of the condensate.

The recent discovery of superfluidity in cold fermion gases opens the
door to a set of fascinating problems in condensed matter physics.
Unlike the bosonic system, there is no counterpart of Gross-Pitaevskii
theory. A new theory which goes beyond BCS and encompasses
BEC in some form or another will have to be developed in concert with
experiment.  \textit{As of this writing, there are four experiments
  where the simple mean field theory discussed in this review is in
  reasonable agreement with the data}.  These include the collective mode
studies over the entire range of accessible magnetic fields
\cite{Tosi,Heiselberg}.  In addition in the unitary regime, RF
spectroscopy-based pairing gap studies \cite{Torma2,heyan}, as well as
density profile \cite{JS5} and thermodynamic studies
\cite{ThermoScience,ChenThermo} all appear to be compatible with this
theory. Interestingly, all of these provide indications for a pseudogap
either directly through the observation of the normal state energy
scales, $T^*$ and $\Delta$, or indirectly, through the observation of
noncondensed pairs.  The material in this Review is viewed as the first
of many steps in a long process.  It is intended to provide continuity
from one community (which has addressed the BCS-BEC crossover scenario,
since the early 1990's) to another.

\acknowledgments

We gratefully acknowledge the help of our many close collaborators over
the years: Jiri Maly, Boldiz\'sar Jank\'o, Ioan Kosztin, Ying-Jer Kao,
Andrew Iyengar, Shina Tan and Yan He.  We also thank our co-authors John
Thomas, Andrey Turlapov and Joe Kinast, as well as Thomas Lemberger,
Brent Boyce, Joshua Milstein, Maria Luisa Chiofalo and Murray Holland.
This work was supported by NSF-MRSEC Grant No.~DMR-0213765 (JS,ST and
KL), NSF Grant No.~DMR0094981 and JHU-TIPAC (QC).

\bibliographystyle{apsrev}

\begin{thebibliography}{116}
\bibitem[{foo()}]{footnoteongapsymmetry}
\bibinfo{note}{To be precise, the gap and pseudogap do not contain a phase and
  thus have $|d_{x^2-y^2}|$ symmetry, while the order parameter has
  $d_{x^2-y^2}$ symmetry.}
\end{thebibliography}

\end{document}